\documentclass[journal, 10pt]{IEEEtran}
\usepackage{cite}
\usepackage{soul}
\usepackage[pdftex]{graphicx}
\usepackage{amsmath}
\usepackage{amsfonts}
\usepackage{amssymb}
\usepackage{color}
\usepackage{float}
\usepackage{color}
\usepackage{epstopdf}
\usepackage{subfigure}
\usepackage[margin=0.45in]{geometry}
\usepackage{verbatim}
\usepackage{subfig}
\usepackage{booktabs}
\usepackage[export]{adjustbox}
\usepackage{multirow}
\usepackage[ruled,vlined]{algorithm2e}

\begin{document}
\newcommand{\mytab}{
	\begin{tabular}{lcr}
		\toprule
		One & Two & Three \\
		\midrule
		$x$ & $y$ & $z$ \\
		1 & 2 & 3 \\
		\bottomrule
	\end{tabular}
}
\title{A Fokker-Planck Approach for Modeling the Stochastic Phenomena in Magnetic and Resistive Random Access Memory Devices}

\author{Debasis~Das and~Xuanyao Fong, \emph{Member, IEEE}
	
	\thanks{The authors are with the Dept. of Elect. \& Comp. Engineering, National University of Singapore,
		Singapore, 117583 (e-mail: \{eledd,kelvin.xy.fong\}@nus.edu.sg).}
}
\maketitle
\begin{abstract}
Embedded non-volatile memory technologies such as resistive random access memory (RRAM) and spin-transfer torque magnetic RAM (STT MRAM) are increasingly being researched for application in neuromorphic computing and hardware accelerators for AI. However, the stochastic write processes in these memory technologies affect their yield and need to be studied alongside process variations, which drastically increase the complexity of yield analysis using the Monte Carlo approach. Therefore, we propose an approach based on the Fokker-Planck equation for modeling the stochastic write processes in STT MRAM and RRAM devices. Moreover, we show that our proposed approach can reproduce the experimental results for both STT-MRAM and RRAM devices.  
\end{abstract}
\begin{IEEEkeywords}
Fokker-Planck equation, Magnetic tunnel junction, MRAM, spin-transfer torque devices, Resistive random access memory, RRAM.
\end{IEEEkeywords}

\section{Introduction}\label{Intro}

\begin{figure}[!b]
	\centering
	\includegraphics[scale=0.15]{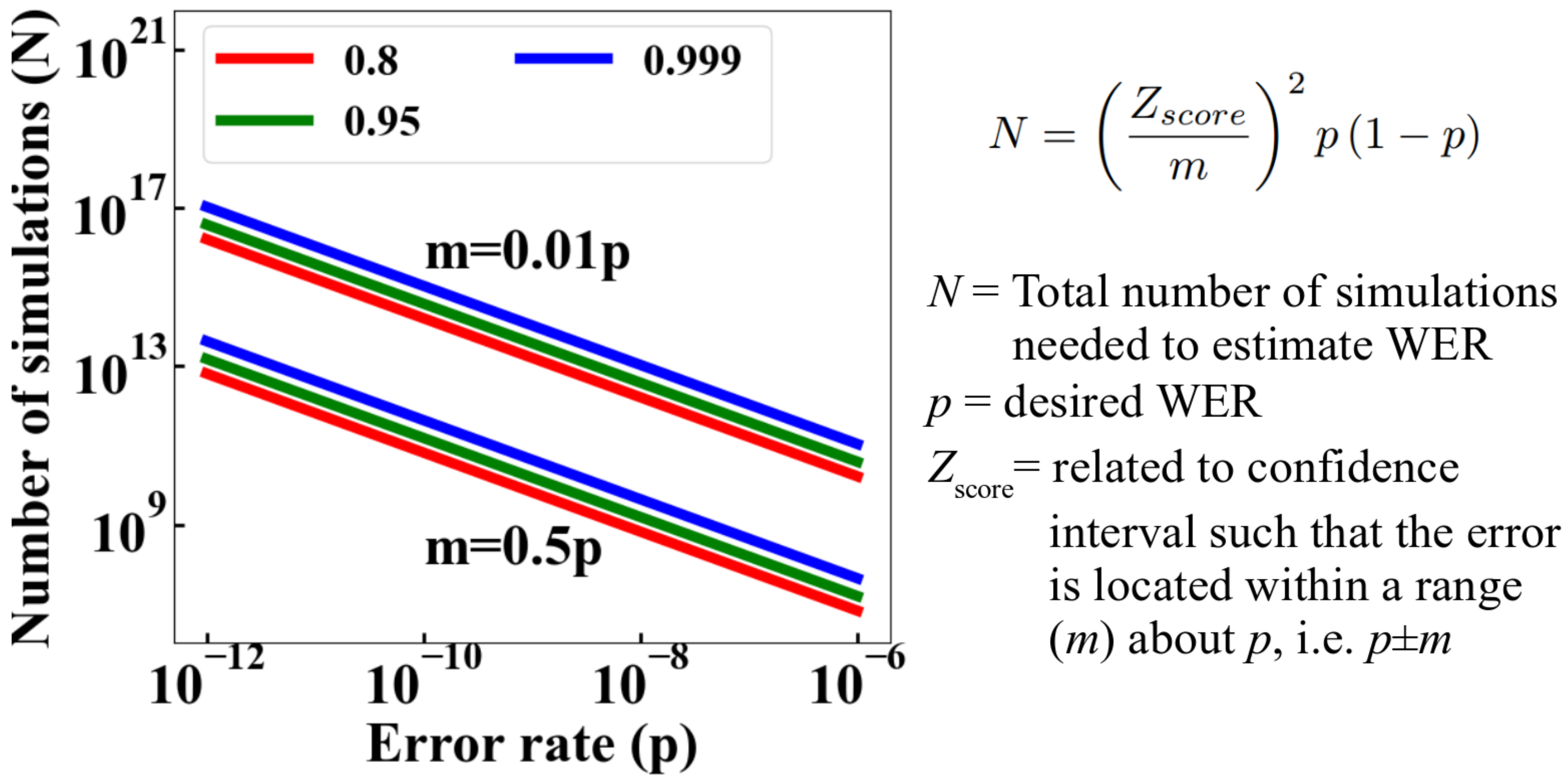}
	\caption{Number of simulations $N$ vs. error rate ($p$) for values of $m=0.5p$ and $m=0.01p$. Different colors in the label denotes the confidence interval.}
	\label{N_vs_error}
\end{figure}

Due to increasing demand for memory-intensive applications such as hardware accelerators for AI, embedded non-volatile memory (eNVM) technologies such as resistive random access memory (RRAM)\cite{wong2012metal} and spin-transfer torque magnetic random access memory (STT MRAM) \cite{fong2016spin} have been widely explored for these applications \cite{giacomin2018robust,chen2019comprrae,jiang2020mint, angizi2019mrima, yan2018celia, pan2018multilevel} due to their comparative advantage (such as non-volatility, near-zero standby leakage power, small cell size and scalability \cite{xu2008bipolar, walczyk2009pulse,hosomi2005novel,ohsawa20121mb, noguchi2014highly, lee2008low, nam2006switching, das2017scaling, tseng2009high}) over other memory technologies. However, the programming mechanisms in RRAM \cite{yu2011stochastic} and STT MRAM \cite{fong2013failure} are inherently stochastic, which introduce write errors that may drastically impact the robustness and reliability of the memory and thus, pose a significant challenge to their widespread adoption. Moreover, the impact of the write error rate (WER) needs to be investigated alongside process variations during yield analysis, which may significantly increase the complexity of conventional methods based on the Monte Carlo approach. Consider the simulation of WER in STT MRAM for example. The WER is estimated by counting the number of failures out of $N$ simulation runs and the probability of obtaining each observation follows a binomial distribution. The number of simulation runs needed to estimate the WER and the accuracy of the estimation are graphed in Figure~\ref{N_vs_error}. Thus, a significant amount of computing resources may be required to perform yield analysis in RRAM and STT MRAM.

Alternatively, it was shown that the WER of STT MRAM can be estimated using an approach based on the Fokker-Planck (FP) equation \cite{xie2016fokker,miriyala2019influence}. In this approach, the distribution of memory cell states during the write process is modeled directly, which significantly reduces the computation effort for estimating the WER. In this paper, we propose a generalized approach for estimating the WER in STT MRAM and steady state analysis of RRAM devices, based on the Fokker-Planck equation. Our approach models the distribution of (i) magnetization state in STT MRAM, and (ii) filament gap lengths in RRAM, to estimate the WER in the corresponding memory device. We further show that our simulation approach is able to reproduce experimentally measured WER in STT MRAM and filamentary RRAM.

The rest of the paper is organized as follows. Our proposed modeling approach using the FP equation is discussed in Section~\ref{sim_frame}. Section~\ref{res_dis} presents our simulation results and discussions. Finally, Section~\ref{conclusion} concludes this paper.

\section{The Proposed Model}\label{sim_frame}

The FP equation, which is written as:
\begin{align}
\begin{split}
\frac{\partial\rho\left(\boldsymbol{x}, t\right)}{\partial t} & = -~div~\left(\rho\left(\boldsymbol{x}, t\right)\boldsymbol{\vartheta}\left(\boldsymbol{x}, t\right) - D \boldsymbol{\nabla}\rho\left(\boldsymbol{x}, t\right)\right) \\
& = -\boldsymbol{\nabla}\cdot \left( \rho\left(\boldsymbol{x}, t\right)\boldsymbol{\vartheta}\left(\boldsymbol{x}, t\right)\right) + D \boldsymbol{\nabla}^2 \rho\left(\boldsymbol{x}, t\right) \label{FP}
\end{split}
\end{align}may be interpreted as a continuity equation that describes the temporal evolution of the probability density function, PDF, of a parameter of interest. $div()=\boldsymbol{\nabla}\cdot$ is the divergence operator, $t$ is time, and the scalar field, $\rho(\boldsymbol{x},t)$, is the PDF of the parameter of interest. If $\boldsymbol{S}$ is the set of all values the parameter of interest can have, then $\boldsymbol{x}\in\boldsymbol{S}$ and $\boldsymbol{S}$ is the state space of $\boldsymbol{x}$. The drift term, $\rho\left(\boldsymbol{x}, t\right)\boldsymbol{\vartheta}\left(\boldsymbol{x}, t\right)$, captures the dynamics of $\rho\left(\boldsymbol{x}, t\right)$ under the influence of deterministic processes, which are modeled using the ``velocity field'', $\boldsymbol{\vartheta}\left(\boldsymbol{x}, t\right)$. The diffusion term, $D\boldsymbol{\nabla}\rho\left(\boldsymbol{x}, t\right)$, models the stochastic nature of $\rho\left(\boldsymbol{x}, t\right)$. In the FP picture, experimentally observed $\rho\left(\boldsymbol{x}, t\right)$ is a result of the competition between the drift and diffusion terms. Thermally-induced stochasticity in $\rho\left(\boldsymbol{x}, t\right)$ may be modeled by considering the diffusion coefficient as a function of temperature (\emph{i.e.} $D=D(T)$ where $T$ is the temperature). To conserve the total probability, Neumann boundary condition need to be assumed. Next, we will discuss how  the FP approach may be used to model the time-dependent dynamics of storage device state distribution in STT MRAM and RRAM.

\subsection{Spin-transfer Torque Magnetic RAM (STT MRAM)}

\begin{figure}[!t]
	\subfigure[]{\includegraphics[scale=0.14]{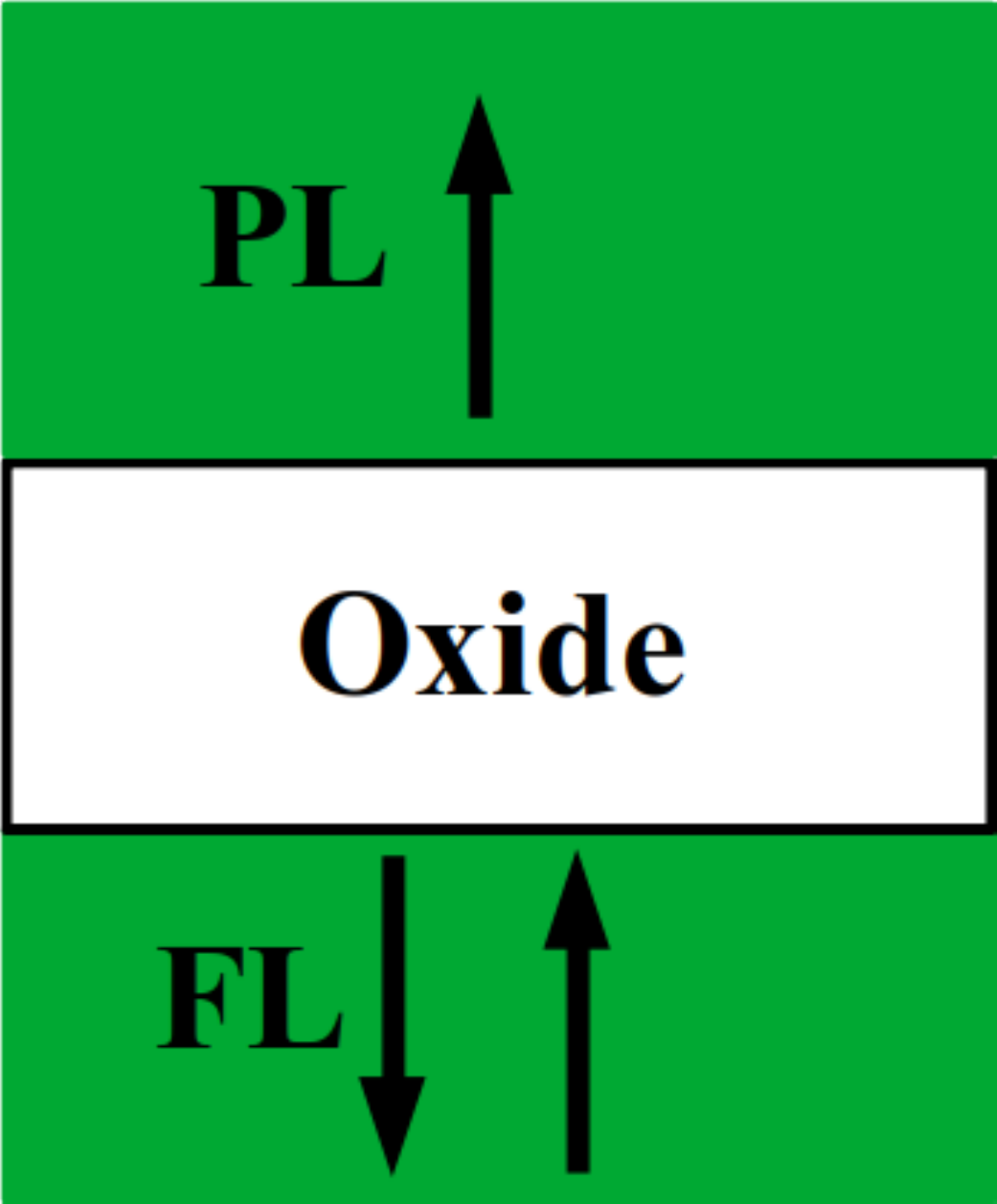}}\label{MTJ_schematic}
	\hfill
	\subfigure[]{\includegraphics[scale=0.12]{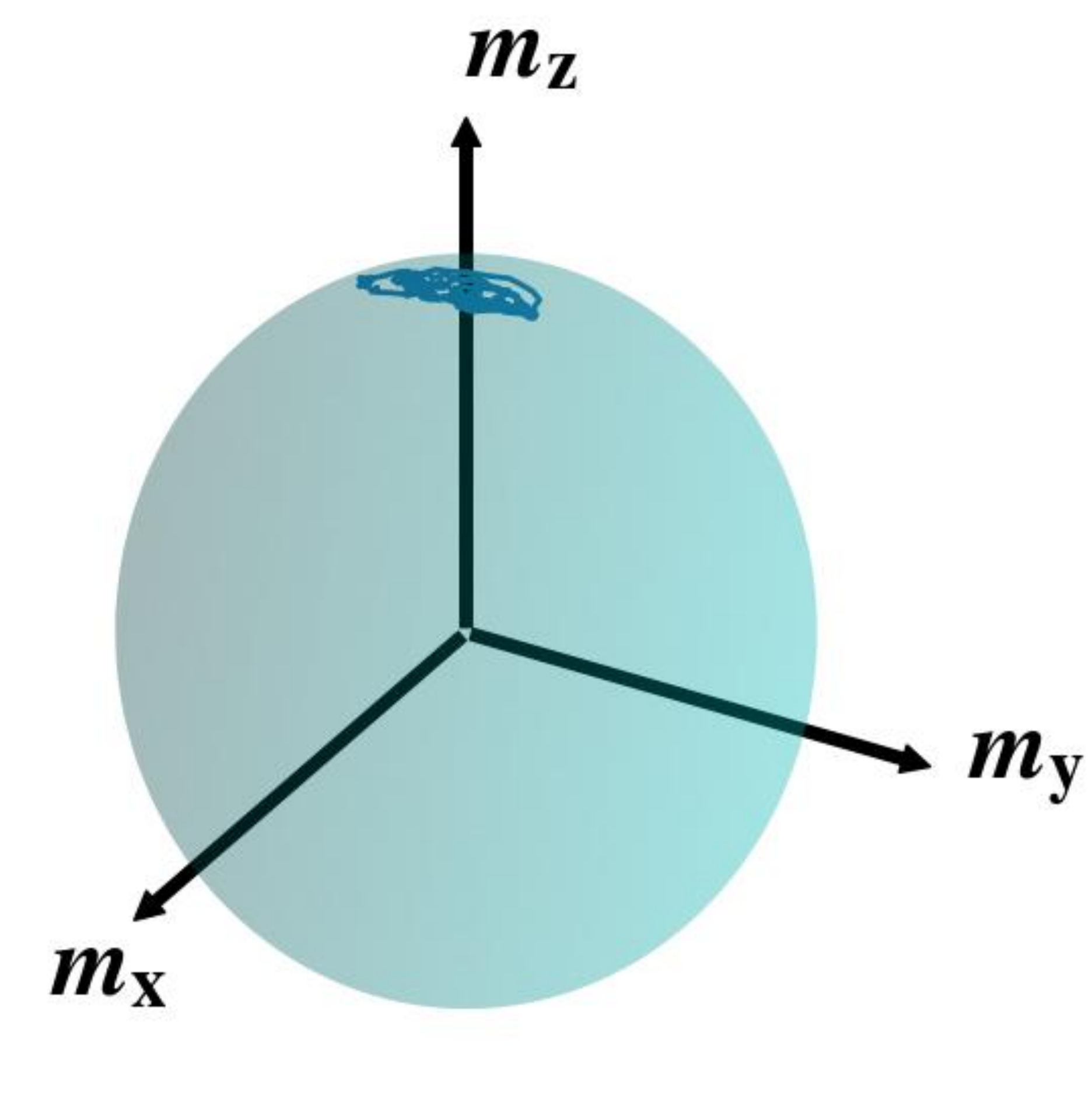}}\label{MTJ_stochastic}
	\hfill
	\subfigure[]{\includegraphics[scale=0.11]{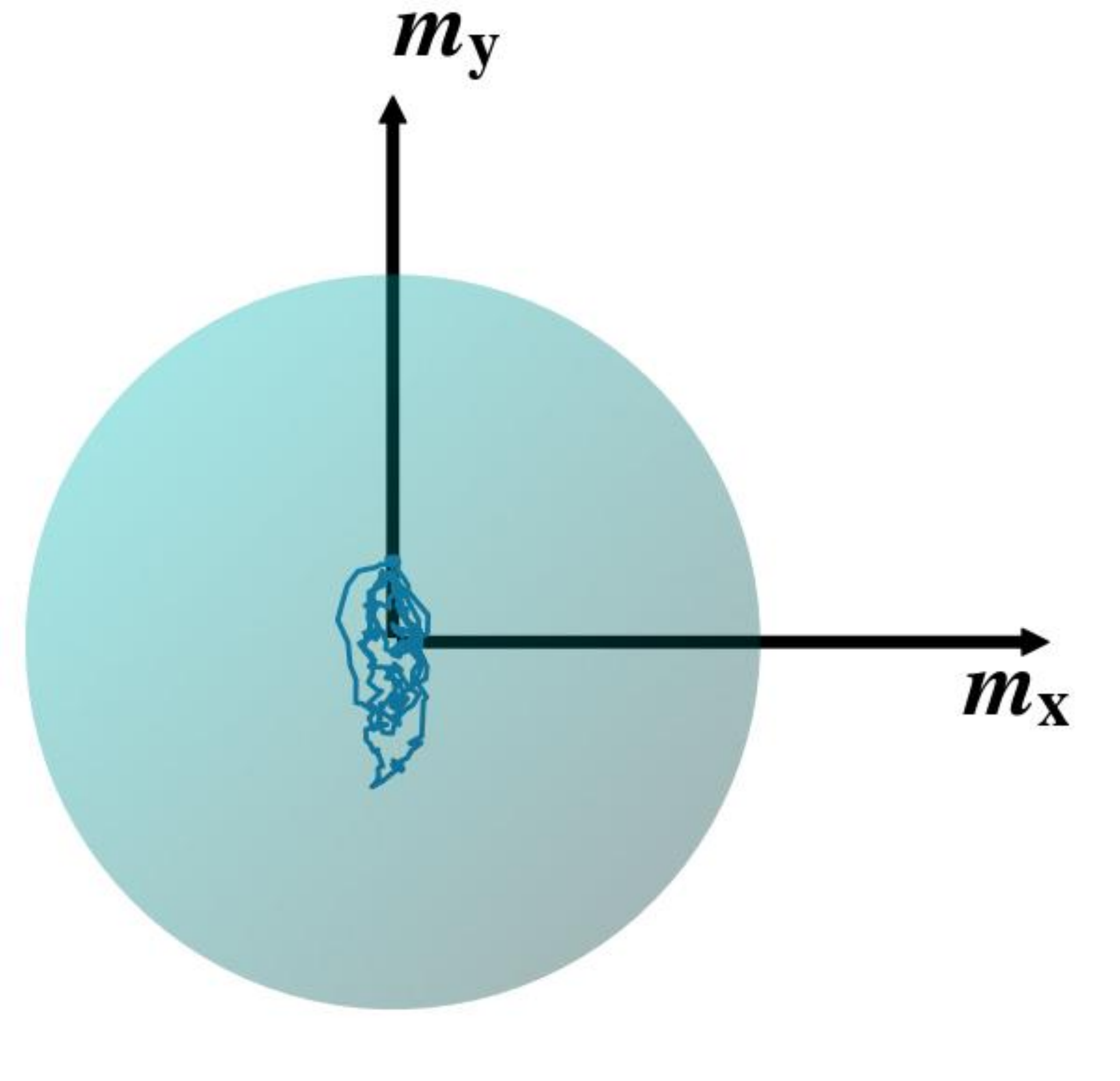}}\label{MTJ_stochastic_top_view}
	\hfill
	\subfigure[]{\includegraphics[scale=0.13]{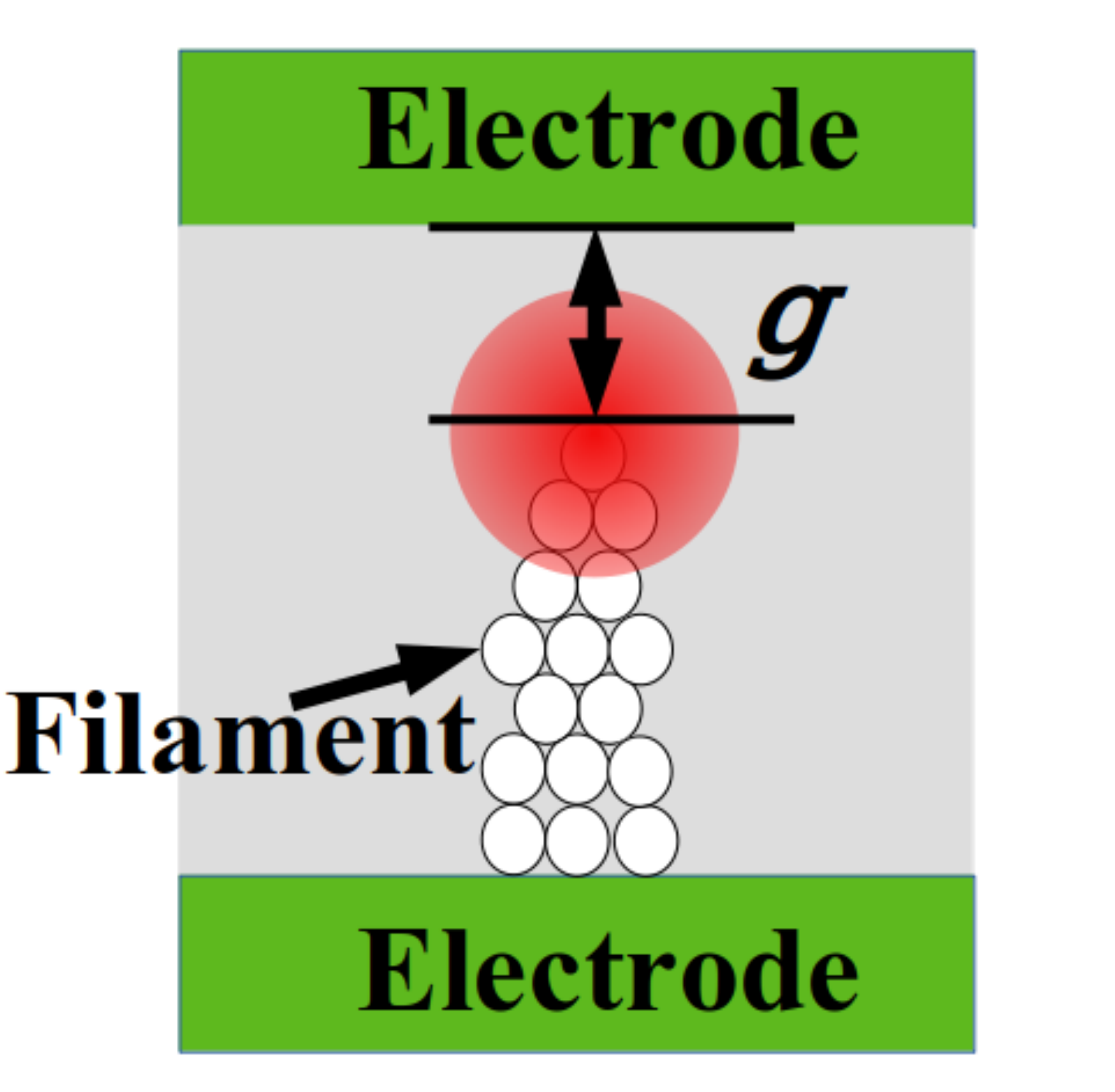}}\label{RRAM_schematic}
		
	\caption{(a) Schematic of MTJ, where PL and FL are separated by an oxide layer. (b) Stochastic nature of the magnetization tip (blue line) on the unit sphere of the state space at zero current. (c) The top-down view of the magnetization tip dynamics over the state space. (d) The schematic of RRAM, where the gap between top of the conductive filament and the opposite electrode is denoted as gap length $g$. The red region at the top of the filament denotes variation in $g$ due to stochasticity.}
	\label{MTJ_RRAM_schematic}	
\end{figure}

In STT MRAM, the storage device is the magnetic tunnel junction (MTJ), which consists of two ferromagnetic (FM) layers sandwiching a tunneling oxide layer (typically MgO) as shown in Fig. \ref{MTJ_RRAM_schematic}(a). The magnetization of one FM layer, called the pinned layer (PL), is fixed whereas the magnetization of the other FM layer, called the free layer (FL), can be changed by current-induced spin torques \cite{albert2000spin}. Also, the MTJ has two stable configurations. In the parallel (P) configuration, the magnetizations of both the PL and the FL point in the same direction. In the antiparallel (AP) configuration, the magnetizations of the PL and the FL point in opposite directions. The MTJ configuration can be electrically sensed using the tunneling magneto-resistance (TMR) effect \cite{fong2013failure}. The magnetization dynamics of the FL is described by the Landau-Lifshitz-Gilbert-Slonczewski (LLGS) equation \cite{slonczewski1996current,xie2016fokker},
\begin{align}
	\begin{split}
	\frac{\left(1+\alpha^2\right)}{\gamma}\frac{\partial\boldsymbol{m}}{\partial t} = & -\mu_{0}\boldsymbol{m} \times \left(\boldsymbol{H}_\text{eff} + \alpha \boldsymbol{m} \times \boldsymbol{H}_\text{eff} \right) \\
	& -\frac{\hbar \epsilon I_\text{MTJ}  }{2 q M_\text{S} \Omega} \left(\boldsymbol{m} \times \boldsymbol{m} \times \boldsymbol{m}_\text{p}\right)
	\end{split}	\label{LLGS}
\end{align}where $\boldsymbol{H}_\text{eff}=\boldsymbol{H}_\text{ua}+\boldsymbol{H}_\text{demag}+\boldsymbol{H}_\text{Zeeman}+\boldsymbol{H}_\text{thermal}$. Here, $\boldsymbol{H}_\text{ua}$, $\boldsymbol{H}_\text{demag}$, $\boldsymbol{H}_\text{Zeeman}$ and $\boldsymbol{H}_\text{thermal}$ are the uniaxial anisotropy field, demagnetization field, the externally applied magnetic field, and the thermal fluctuation field, respectively. In Eq.~(\ref{LLGS}), $\hbar$, $q$, $\alpha$, $\mu_0$, and $\gamma$ are the reduced Planck constant, the electronic charge, the damping factor, the permeability of free space, and the gyromagnetic ratio, respectively. $\boldsymbol{m}$ and $\boldsymbol{m}_\text{p}$ are the magnetization vectors describing the magnetization of FL and PL, respectively. The effective spin polarization factor is calculated as \cite{miriyala2019influence}
\begin{equation}
	\epsilon = \frac{Pol~ \Lambda}{\left(\Lambda^2 + 1\right)+\left(\Lambda^2 - 1\right)\left(\boldsymbol{m}\cdot\boldsymbol{m}_p\right)}
\end{equation}
where, $Pol$ and $\Lambda$ are the polarization of FM and the polarization asymmetry parameter respectively.

For data storage applications, the MTJ is engineered with two stable states, which are separated by an energy barrier, $E_\text{B} = \Delta k_\text{B}T$,  along the uniaxial anisotropy direction. $\Delta$ is called as the thermal stability factor and is given by \cite{butler2012switching}
\begin{equation}
\Delta = \frac{\mu_0 H_\text{eff} M_\text{S} \Omega}{2 k_\text{B} T}
\end{equation}where $M_\text{S}$ is the saturation magnetization, $\Omega$ is the volume of the free layer, $k_\text{B}$ is the Boltzmann constant and $T$ is the temperature.

In Eq.~(\ref{LLGS}), $I_\text{MTJ}$, denotes the current flowing through the MTJ and is given by $I_\text{MTJ} = V_\text{MTJ}/R_\text{MTJ}$, where $V_\text{MTJ}$ is the voltage applied between PL and FL of the MTJ and $R_\text{MTJ}$ is the resistance of the MTJ, given by \cite{miriyala2019influence}
\begin{equation}
	R_\text{MTJ} = R_\text{P} + \frac{\left(R_\text{AP} - R_\text{P}\right)\left(1-\boldsymbol{m}\cdot\boldsymbol{m}_\text{p}\right)}{2\left(1+\left(\frac{V_\text{MTJ}}{V_\text{bias}}\right)^2\right)} \label{R_MTJ}
\end{equation}
$R_\text{P}$ and $R_\text{AP}$ are the resistances of the MTJ in the P and AP configurations, respectively.

As mentioned earlier, deterministic processes are modeled using the ``velocity field'', $\boldsymbol{\vartheta}$, in the FP equation. Thermally-induced fluctuations in $\boldsymbol{m}$ are modeled using $\boldsymbol{H}_\text{thermal}$ in Eq.~(\ref{LLGS}). If $\boldsymbol{H}_\text{thermal}$ is excluded, Eq.~(\ref{LLGS}) captures the deterministic magnetization dynamics of $\boldsymbol{m}$. Hence, the LLGS equation may be used as $\boldsymbol{\vartheta}$ in the FP equation:
\begin{align}
\begin{split}
\boldsymbol{\vartheta}\left(\boldsymbol{m}, t\right) = & -\mu_0 \gamma' \boldsymbol{m} \times \left(\boldsymbol{H}_\text{eff} + \alpha \boldsymbol{m} \times \boldsymbol{H}_\text{eff} \right) \\
	& -\frac{\hbar \gamma' \epsilon I_\text{MTJ}  }{2 q M_\text{S} \Omega} \left(\boldsymbol{m} \times \boldsymbol{m} \times \boldsymbol{m}_\text{p}\right) \label{LLGS_velocity}
	\end{split}
\end{align}where $\gamma'=\gamma/\left(1+\alpha^{2}\right)$, and $\boldsymbol{H}_\text{thermal}$ is neglected in $\boldsymbol{H}_\text{eff}$ of Eq.~(\ref{LLGS_velocity}). The state space of $\boldsymbol{m}$ is the surface of the unit sphere centered about the origin. If $\boldsymbol{H}_\text{thermal}$ is included in the $\boldsymbol{H}_\text{eff}$, the tip of $\boldsymbol{m}$ vibrates around the mean point, which is shown in Fig.~\ref{MTJ_RRAM_schematic} (b) and (c).

It was shown in \cite{Nigam2011e} that $\rho(\boldsymbol{m})$ at thermal equilibrium for a PMA-based MTJ is\begin{equation}
P\left(\boldsymbol{m}\cdot\boldsymbol{m}_\text{p}\right) = \frac{2 H_\text{eff} M_\text{S}}{k_\text{B} T}exp\left(-\frac{H_\text{eff} M_\text{S}}{k_\text{B} T}(1-\left(\boldsymbol{m}\cdot\boldsymbol{m}_\text{p}\right)^{2})\right)
\end{equation}Hence, the diffusion coefficient, $D$, for the FP model of the PMA-based MTJ can be calculated as,
\begin{equation}
D = \frac{\alpha \gamma k_\text{B} T}{\left(1+\alpha^2\right)M_\text{S} \Omega}
\end{equation}

Note that phenomena such as spin-orbit torque (SOT) \cite{liu2009reduction} and voltage-controlled magnetic anisotropy (VCMA) \cite{kang2017modeling} can be captured by either adding additional torque terms to Eq.~(\ref{LLGS}) or by adding a corresponding field term to the effective magnetic field. Consequently, $\boldsymbol{\vartheta}(\boldsymbol{m}, t)$ may be modified accordingly to capture effects of SOT and VCMA in the FP model.

\subsection{Filamentary Resistive RAM (RRAM)}

The analog resistance of filamentary RRAM devices arises due to varying gap lengths between the conducting filament and the electrode \cite{jiang2014verilog} as shown in Fig.~\ref{MTJ_RRAM_schematic}(d). Application of an electric field between the electrodes of the filamentary RRAM device causes the filament to grow or shrink due to drift of the atomic species constituting the filament. The exact placement of the atomic species in the filament cannot be precisely controlled and gives rise to the stochastic nature of the gap length, which manifests as the variation of resistance (both cycle-to-cycle and device-to-device) in the measured current-voltage ($I$-$V$) characteristics of the filamentary RRAM device.

To model the filamentary RRAM device using the FP equation, we propose to model $\rho\left(g,t\right)$, which is  the PDF of $g$, the gap length in the filamentary RRAM device. For convenience, we propose that the ``velocity field'', $\boldsymbol{\vartheta}\left(g,t\right)$ may be modeled using Ginzburg-Landau theory \cite{ouyang2004simulation} and the diffusion constant is a homogeneous constant in the filamentary RRAM device. In our proposed method, a free energy function, $f_\text{free}(g, t)$, is used to directly capture the distribution of $g$, where 
\begin{equation}
	\boldsymbol{\vartheta}\left(g, t\right) = -\kappa\frac{\partial f_\text{free}\left(g, t\right)}{\partial g}
\end{equation}Consequently, $f_\text{free}\left(g, t\right)$ behaves like a potential well and the interaction between the drift and diffusion terms in the FP equation gives rise to the observed distribution of $g$.

As the scope of this paper is to motivate the use of the FP approach to study WER in STT MRAM and filamentary RRAM, further development of the theory behind the FP model for filamentary RRAM will be left as future work to be pursued. As will be discussed in Section~\ref{RRAM_res}, in this work, we will show that our proposed FP model can accurately reproduce the distributions of experimentally measured filamentary RRAM resistances.

\section{Results and Discussion} \label{res_dis}

In this section, we discuss the FP simulation results for the STT MRAM and RRAM. Our FP equation solver is implemented in the Python-based `FEniCS' library \cite{langtangen2016solving}, which solves the partial differential equation (PDE) using the finite element method.

\subsection{STT MRAM}\label{MTJ_results}

We first investigate the relationship between WER and the voltage applied across the MTJ. In an MTJ, $\boldsymbol{m}$ can be either in-plane due to in-plane magnetic anisotropy (IMA) or perpendicular-to-plane due to perpendicular magnetic anisotropy (PMA) \cite{fong2016spin, huai2008spin}. MTJs with PMA are preferred over those with IMA due to their energy efficiency and high thermal stability \cite{mangin2006current, nakayama2008spin,ikeda2010perpendicular} and thus, we focus on PMA-based MTJs in this work. For the PMA-based MTJ, $H_\text{ua}$ and $H_\text{demag}$ act opposite to each other \cite{liu2009reduction}. The parameter values for the PMA-based MTJ in our simulation were extracted from Ref \cite{nowak2016dependence} and \cite{xie2016fokker}, where the FL is assumed to be cylindrical shape with 40~nm diameter circular cross-section and 1~nm thickness. 

Our simulations are initialized as follows. A 2~T external magnetic field is initially applied along the $+z$-direction for 10~ns and the FL is allowed to relax for a further 5~ns with the external magnetic field reduced to 0~T. Since the state space of $\boldsymbol{m}$ is the surface of the unit sphere centered about the origin, to calculate the WER, we partition the unit sphere into the upper hemisphere ($m_\text{z}>0$) and lower hemisphere ($m_\text{z}<0$). While considering the switching of the MTJ from P to AP, we considered $\boldsymbol{m}_\text{p}=+\widehat{z}$ ($\widehat{z}$ is the unit vector pointing in the $+z$-direction) whereas $\boldsymbol{m}_\text{p}=-\widehat{z}$ when considering the MTJ to be switching from AP to P. Therefore, the WER is estimated by calculating the probability of finding $m_\text{z}\geq 0$.

\begin{figure}[!t]
	\includegraphics[scale=0.2]{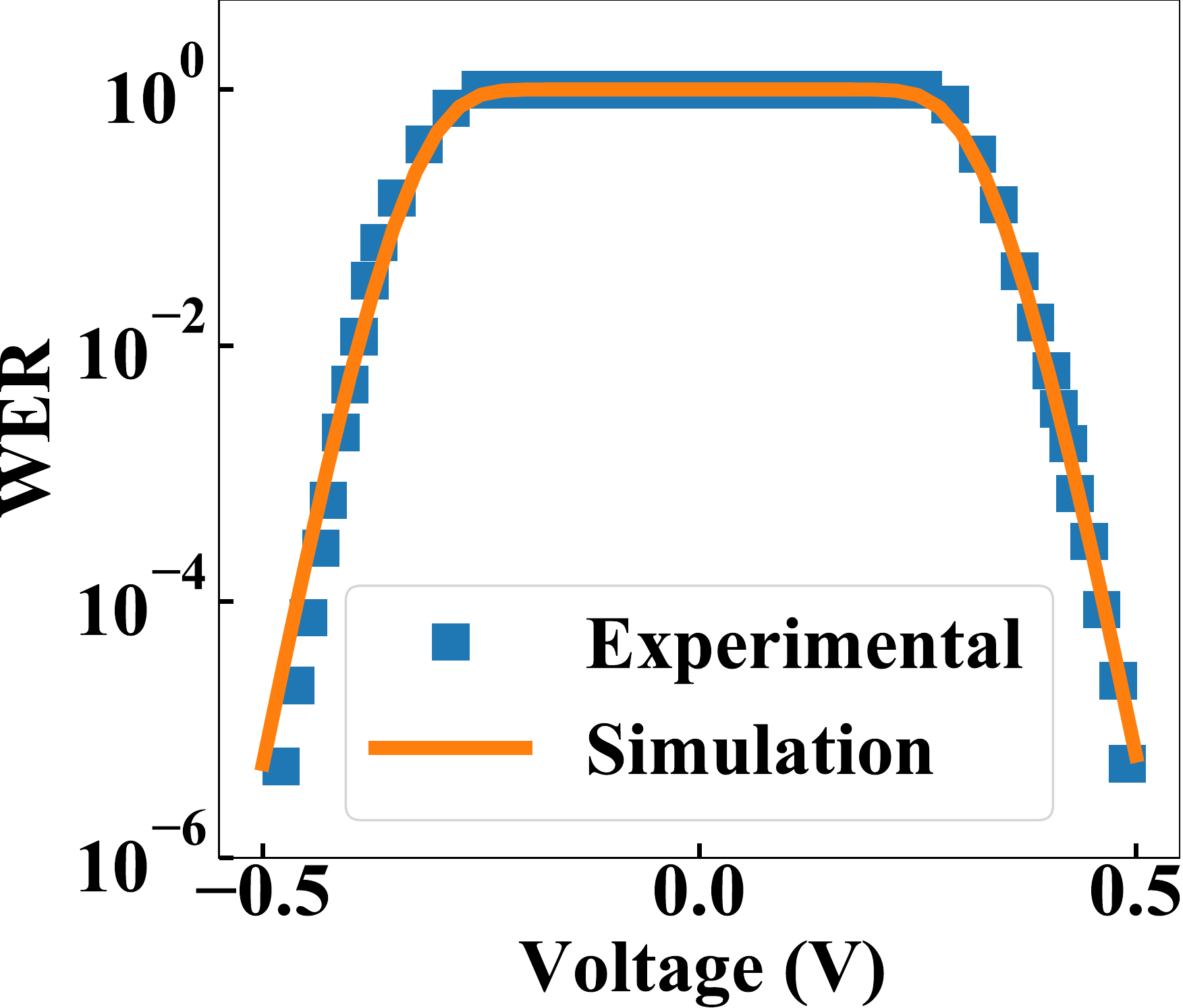}
	\hfill
	\includegraphics[scale=0.17]{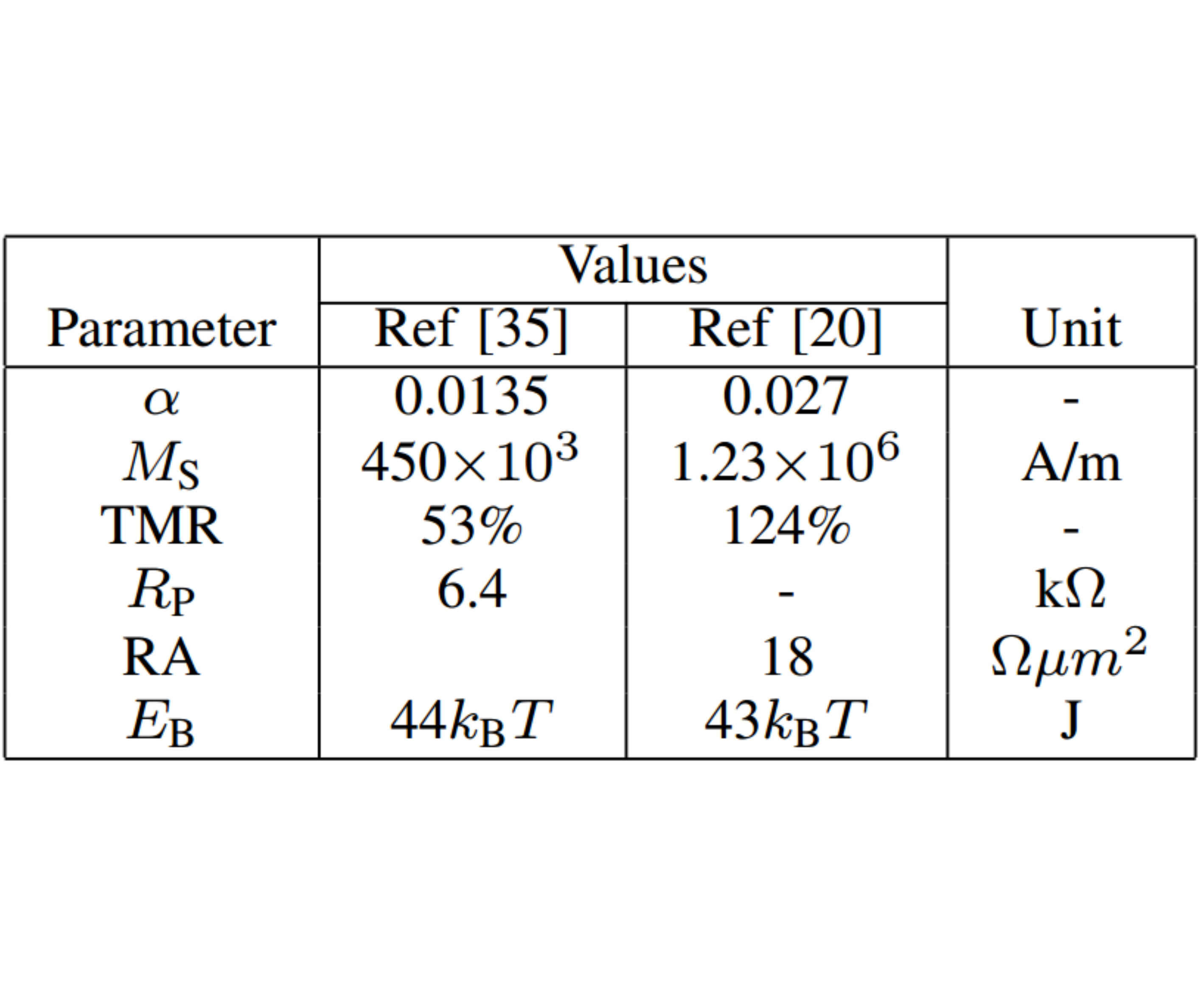}
	\caption{Plot of WER with voltage for the FP simulation. Benchmarking the simulation result with extracted experimental data \cite{nowak2016dependence}. }
	\label{WER_expt_vs_voltage}
\end{figure}

Fig.~\ref{WER_expt_vs_voltage} shows the variation of WER with the voltage applied (for 10~ns pulse width) across the MTJ along with the simulation parameters used. From this plot, it can be seen that our simulation results matches very closely with the experimental result reported in \cite{nowak2016dependence} and demonstrates successful calibration of our simulation framework to experimental measurements.

\begin{figure}[!t]
	\subfigure[]{\includegraphics[scale=0.24]{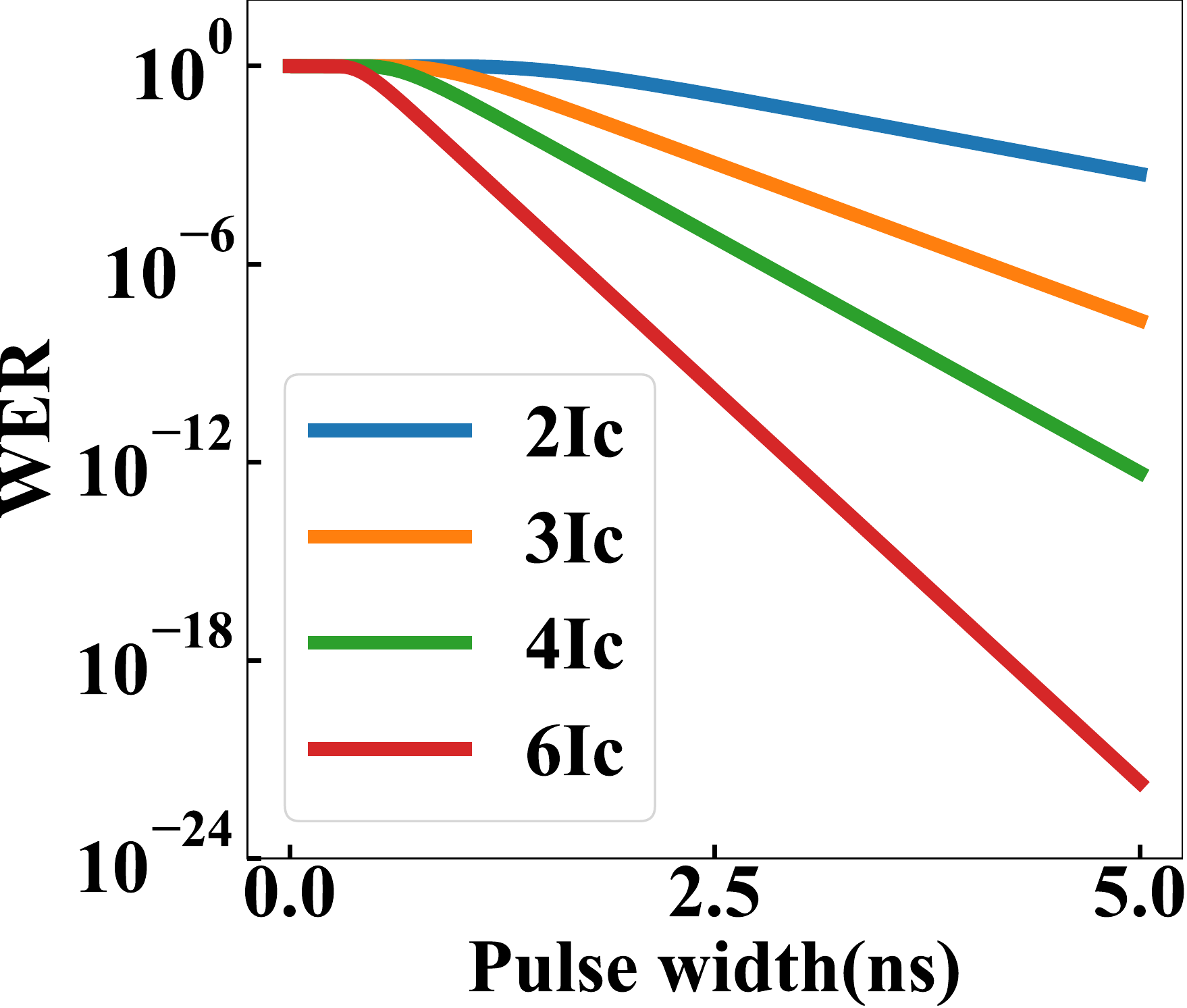}}\label{WER_plot_const_current}
	\hfill
	\subfigure[]{\includegraphics[scale=0.24]{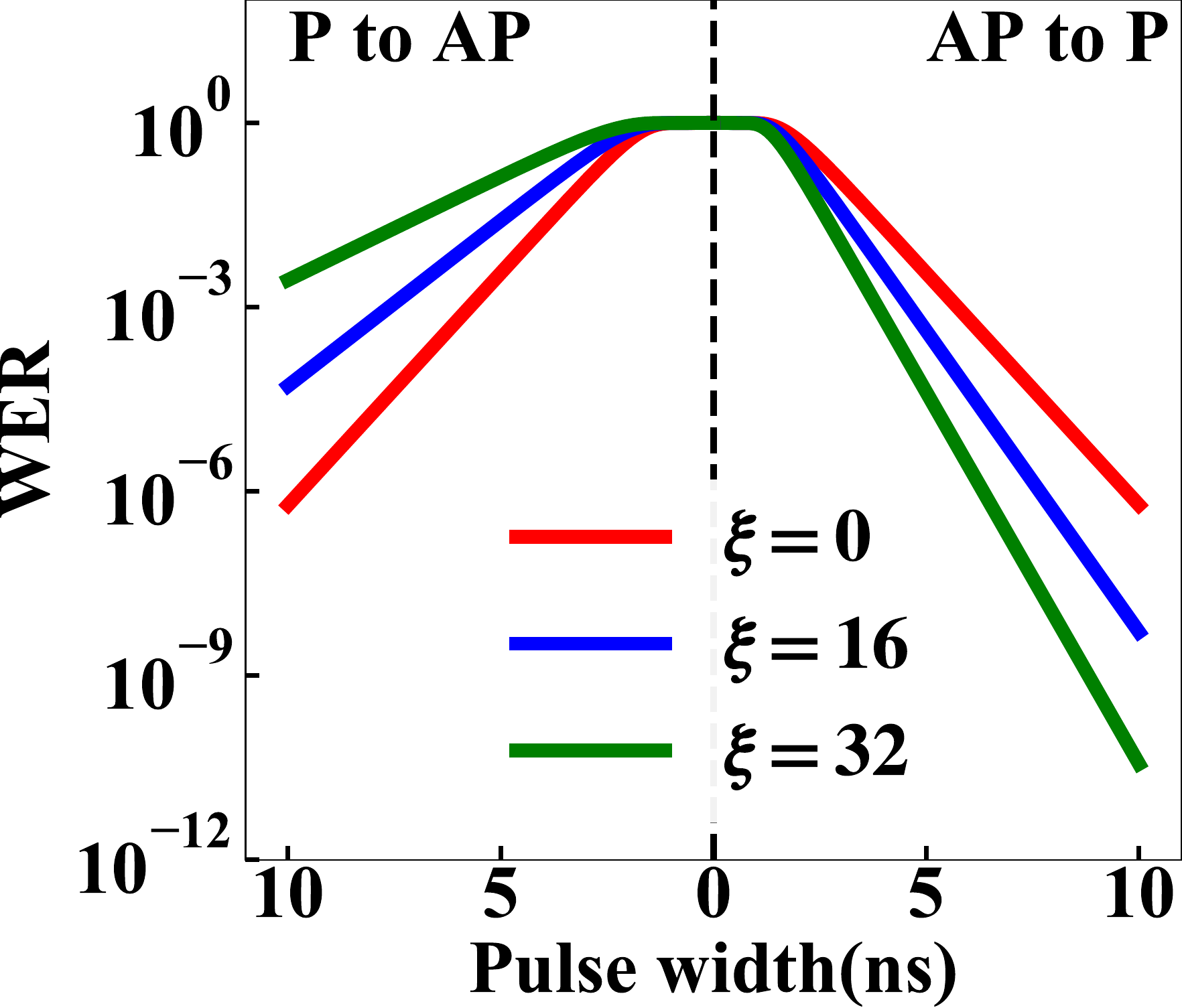}}\label{WER_plot_voltage}
	\caption{Plot of WER as function of pulse width for (a) different charge current, in multiple of $I_C$ ; and (b) voltage without and with VCMA. The red line shows WER for without VCMA ($\xi=0$) and blue and green line shows WER for $\xi$=16 and 32 $\mathrm{fJ/V.m}$ respectively.}
	\label{WER_curr_V}
\end{figure}

Next, we plot WER as a function of pulse width for different charge currents passed through the MTJ as shown in Fig.~\ref{WER_curr_V}(a). For this simulation, we have used the parameters of Ref \cite{xie2016fokker}. When a voltage is applied across the MTJ, a charge current flows through the PL and generates a spin current that eventually switches the nanomagnet of the FL if it exceeds a critical value calculated as
\begin{equation}
	I_\text{c}=\frac{2 q \alpha}{\epsilon \hbar}\mu_0 H_\text{k} M_\text{S} \Omega
\end{equation} 
From Fig.~\ref{WER_curr_V}(a), it can be seen that, for any fixed charge current, as the pulse width increases, WER decreases because more polarized spins will be available within the time window to switch the FL nanomagnet. It also can be seen that, for a fixed pulse width, WER decreases with increasing the charge current. As the charge current magnitude increases, the polarized spin current increases as well and leads to better switching and hence, WER decreases.

The previous simulations considered a constant current $I_\text{MTJ}$ in the last term of R.H.S. of Eq. (\ref{LLGS}). In circuits, it may be costly to design a current source to perform the write operation on the MTJ, and the $I_\text{MTJ}$  generated by applying $V_\text{MTJ}$ depends on $R_\text{MTJ}$. Note that $R_{\text{MTJ}}$ depends on the term $m\cdot m_\text{p}$ in Eq. (\ref{R_MTJ}). Thus, we are able to investigate the variation in the WER by considering the application of $V_\text{MTJ}$ while accounting for $R_\text{MTJ}$. The WER variation with the pulse width of $\pm$2~V is simulated for both AP to P and P to AP respectively, as shown by the red line in Fig.~\ref{WER_curr_V}(b). The plot of WER vs. pulse width of constant voltage shows similar characteristics as the earlier results, where WER decreases with increasing pulse width. The reason can be explained similarly, that with the increase of pulse width, more spin-polarized electrons enter into FL to switch the nanomagnet, which leads to lower WER. Also, for 10~ns pulse width, WER~$\approx~10^{-6}$.

\begin{figure*}[!h]
	\subfigure[]{\includegraphics[scale=0.22]{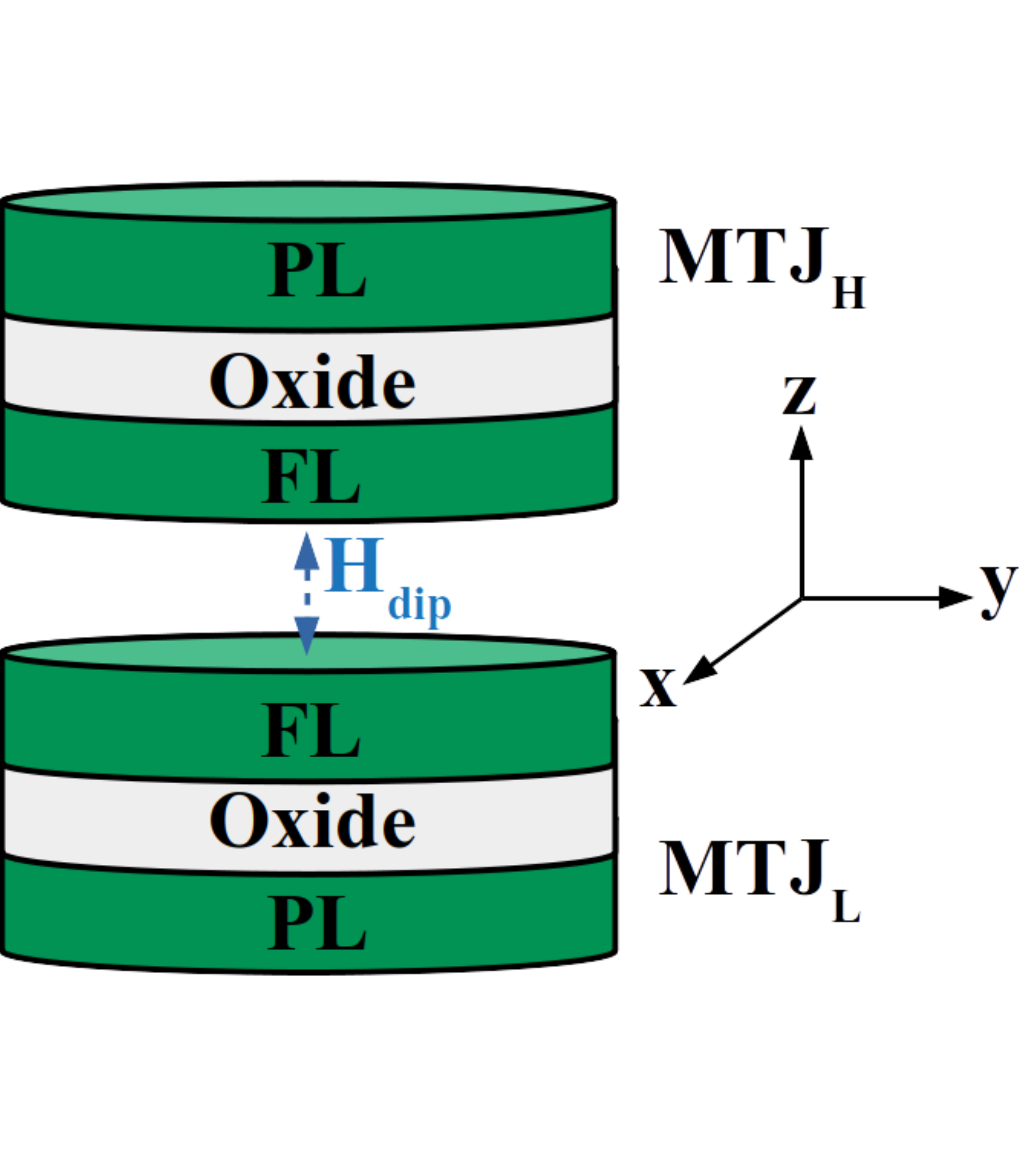}}
	\hfill
	\subfigure[]{\includegraphics[scale=0.25]{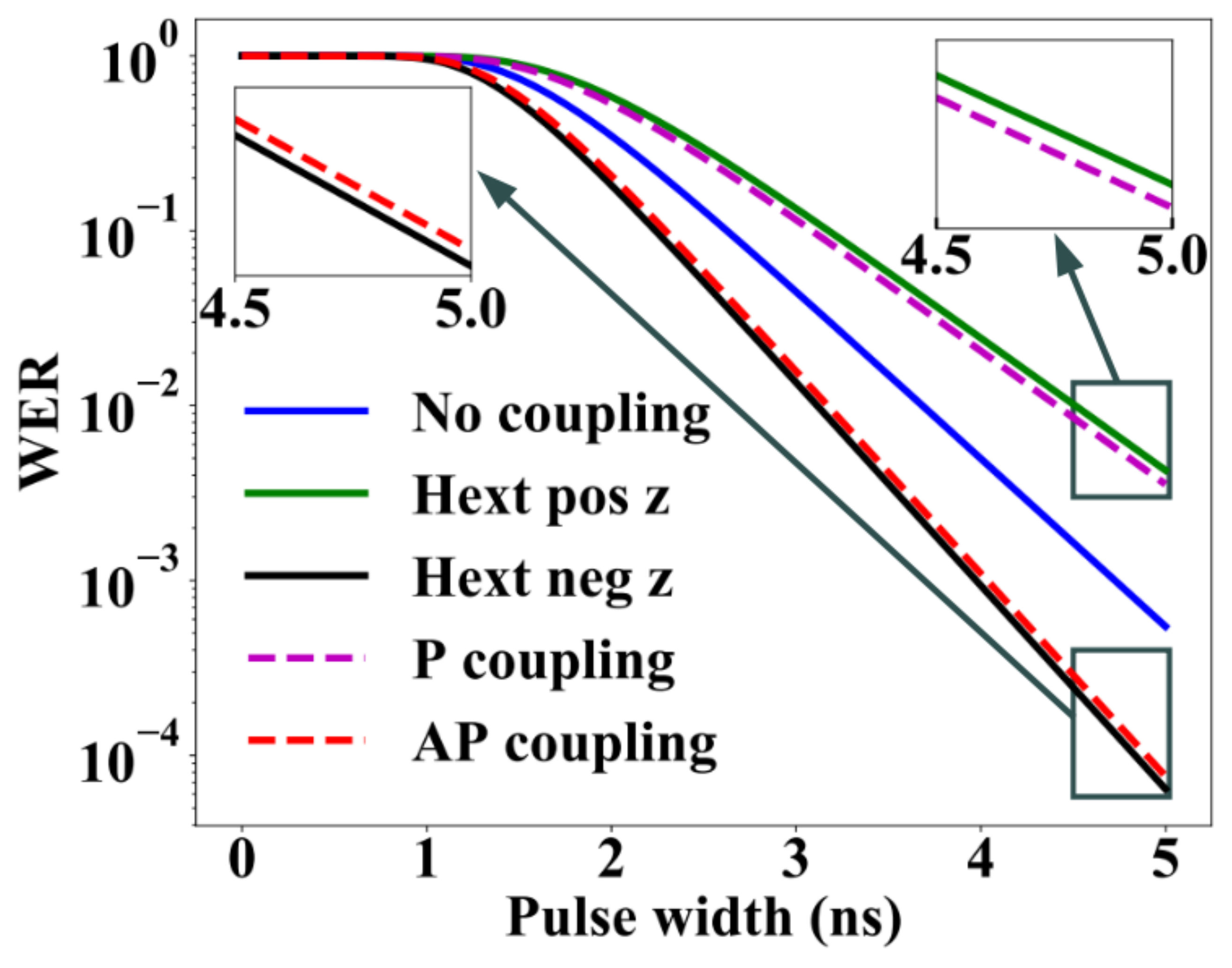}}
	\hfill
	\subfigure[]{\includegraphics[scale=0.276]{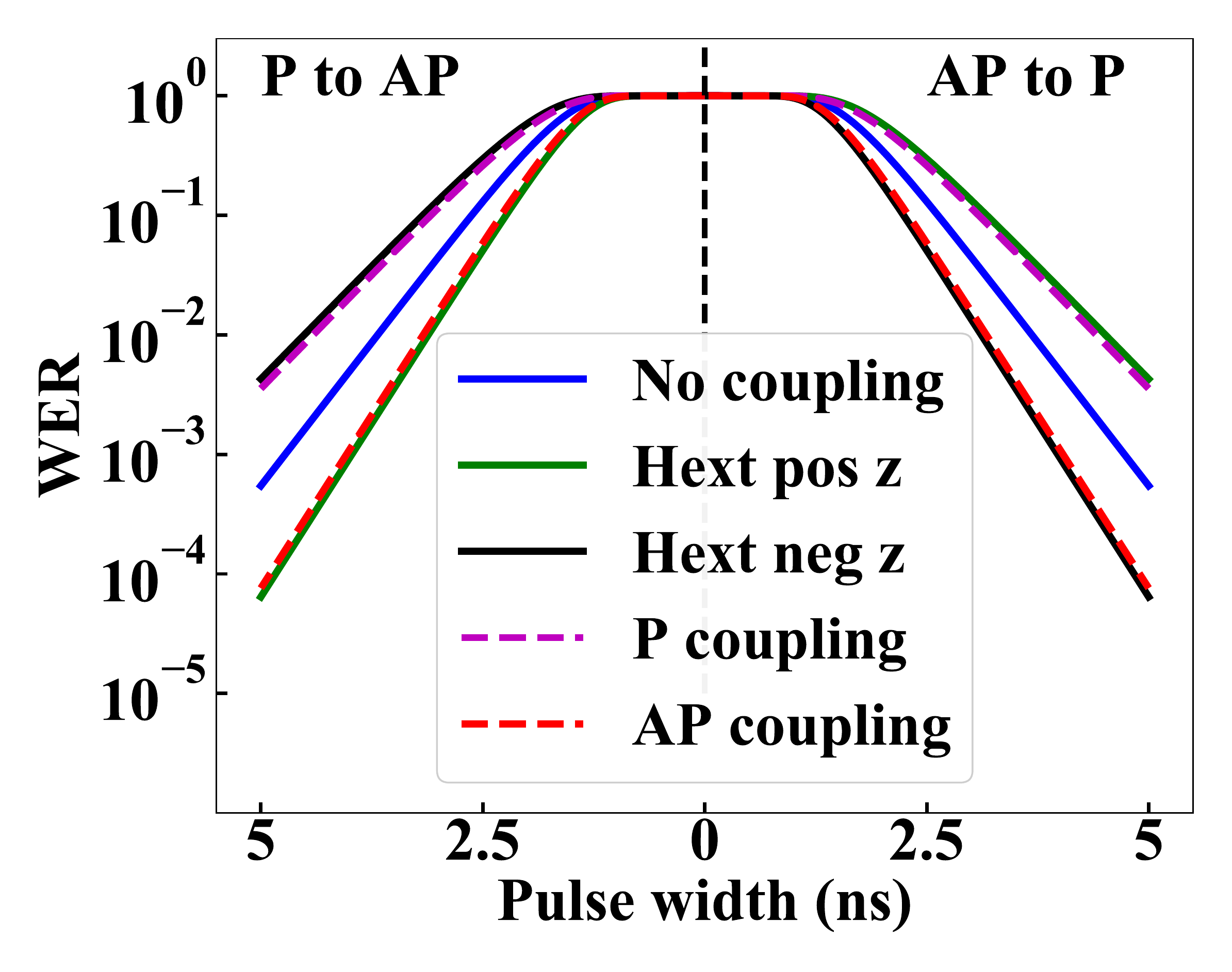}}
	\caption{(a) Schematics of two electrically separated nanomagnets that are coupled through dipolar field $H_{dip}$, (b) plot of WER as a function of pulsewidth for no dipolar coupling, P coupling, AP coupling, and the magnetic field applied along positive and negative $z$-direction. The inset on the left side shows a zoomed-in view for a duration of 4.5-5 $ns$ of the AFM coupling and the magnetic field applied along negative $z$-direction. On the right side, the inset shows a zoomed-in view of the FM coupling and the magnetic field applied along positive $z$-direction for the same time duration, (c) plot of WER vs pulsewidth for for no dipolar coupling, P coupling, AP coupling, and the magnetic field applied along positive and negative $z$-direction for P-to-AP and AP-to-P switching of the nanomagnet. }
	\label{dipolar}
\end{figure*}

We now study the WER if the MTJ is engineered with the voltage-controlled magnetic anisotropy (VCMA) effect. The uniaxial field is given by
\begin{equation}
	H_{ua}=\frac{2 K_\text{u}}{\mu_0 M_\text{S}} 
\end{equation}
where $K_\text{u}$ denotes the uniaxial anisotropy constant. $K_\text{u}$ can be modulated by applying a voltage across the MTJ \cite{weisheit2007electric,amiri2015electric} which is given by,
\begin{equation}
	K_\text{VCMA}=\frac{\xi V_\text{MTJ}}{t_\text{OX} t_\text{FL}}
	\label{Kvcma}
\end{equation}
where $\xi$ is VCMA coefficient, $t_\text{OX}$ and $t_\text{FL}$ are the thicknesses of the oxide and FL, and the anisotropy constant is modified as $K_\text{u}(V_\text{MTJ}) = K_\text{u}(V_\text{MTJ}=0)-K_\text{VCMA}$. The effect of VCMA coefficient is shown in Fig. \ref{WER_curr_V}(b) for $\xi$ = 16 and 32~$\mathrm{fJ/V.m}$ for both AP to P and P to AP switching. For AP to P switching, $V_\text{MTJ}=2~\text{V}$ is applied. From Eq.~(\ref{Kvcma}), $K_\text{VCMA}$ increases with $\xi$ and $K_\text{u}$ and hence, $E_\text{B}$, are reduced, which makes the switching easier. From the plot, it can be seen that VCMA effect improves the WER significantly, down to $\sim 10^{-9}$ and $\sim 10^{-11}$ for $\xi=16$ and 32~$\mathrm{fJ/V.m}$, respectively. On the other hand, for P to AP switching, $V_\text{MTJ}=-2~\text{V}$ is applied. Thus, with the increase of $\xi$, $K_\text{VCMA}$ increases $K_\text{u}(V_\text{MTJ})$ (and hence, $E_\text{B}$), which makes it more difficult to switch the MTJ. This is clearly demonstrated by the increase in WER with $\xi$ in Fig.~\ref{WER_curr_V}(b).

Till now, we have considered a monodomain FL of the MTJ, which is sufficient for studying MRAMs based on STT, SOT, and VCMA switching mechanism. On the other hand, an emerging class of magnetic devices \cite{atulasimha2010bennett,sivasubramani2020nanomagnetic} use coupled (e.g., dipolar coupling) monodomains in the device to achieve logic and memory functionality. Such devices may also be affected by thermally-induced switching errors. To model a monodomain that is dipole coupled to another monodomain in our proposed FP based approach, the effect of the dipolar field needs to be modeled. We first consider a pair of monodomains which are separated electrically but coupled to each other either ferromagnetically (FM) or antiferromagnetically (AFM) through a dipolar field. Moreover, consider if the magnitude of the dipolar field is of $H_{dip}=0.2H_{ua}$. Note that this field can be calculated using the dipolar field energy \cite{mccray2020electrically} or through micromagnetics software such OOMMF \cite{OOMMF}. If we consider the magnetizations of the pair of magnets to be always aligned parallel or anti-parallel to each other, the dipolar field becomes dependent on the state-space of our FP approach. For two dipolar coupled magnets ($\mathrm{FL_L}$ and $\mathrm{FL_H}$) that are separated along the direction of uniaxial anisotropy ($z$-direction) as shown in Fig. \ref{dipolar}(a). Therefore, the dipole field is calculated as $\boldsymbol{H}_{dip}=\pm H_{dip}\left(\widehat{z}\left(\boldsymbol{m}\cdot\widehat{z}\right)-\boldsymbol{m}\right)$, where `+'(`-') symbol is for when both magnets are always parallel(anti-parallel) to each other. Now, if a current  $I=2I_c$ with $m_p=-\widehat{z}$ is driven through $FL_L$ for 5 $ns$, the WER is shown in Fig. \ref{dipolar}(b). The blue curve in Fig. \ref{dipolar}(b) (identical to the blue line shown in Fig. \ref{WER_curr_V}(a)) shows the WER in the absence of the dipolar coupling. The WER if both nanomagnets are always parallel to each other is shown by the magenta dashed line in Fig. \ref{dipolar}(b). The WER has increased because the dipolar field increases the energy barrier that needs to be overcome to switch $FL_L$. On the other hand, when the nanomagnets are always anti-parallel to each other, the WER is the red dashed line in Fig. \ref{dipolar}(b). In this case, the dipolar field reduces the enrgy barrier that needs to be overcome to switch $FL_L$ and results in lower WER as compared to the previous two cases. 

To see how this dipolar field acts differently than a simple magnetic field, we examine the model by applying a constant magnetic field with magnitude $H_{dip}$. For comparing it with the P coupling, this magnetic field points along the positive $z$-direction, which is also the initial magnetization direction of the nanomagnet. The WER response of the nanomagnet with the constant magnetic field is similar to the P coupling condition but with a slightly higher WER (as shown by the green continuous line). A zoomed-in view in the inset at the right side of the figure shows this difference. Similarly, for the AP coupling case, a constant magnetic field with magnitude $H_{dip}$ pointing along the negative $z$-direction (i.e. opposite to the initial magnetization) is considered. A similar response of WER is observed but the WER is lower for the constant magnetic field case compared to the AP coupling case (black continuous line). This difference is shown clearly in a zoomed-in view in the inset at the left-side of the figure. For both cases, the difference in WER between the constant magnetic field case and the coupled nanomagnet case is due to the fact that the constant magnetic field points in a fixed direction but the dipolar field of the coupled nanomagnet case is not in a fixed direction. However the effect of the fields on the energy barrier that the nanomagnets need to overcome during switching is similar and hence, the WER are very close to each other. This is clearly indicated by Fig \ref{dipolar}(c), where we have studied the switching from P configuration to AP as well as the reverse. Moreover, due to the fixed direction of the constant magnetic field, switching in one direction is favored but is opposed in the opposite direction. This is observed as the asymmetry in the WER curves for the constant magnetic field cases.

\subsection{RRAM}\label{RRAM_res}

\begin{figure}[!h]
	\centering
	\includegraphics[scale=0.3]{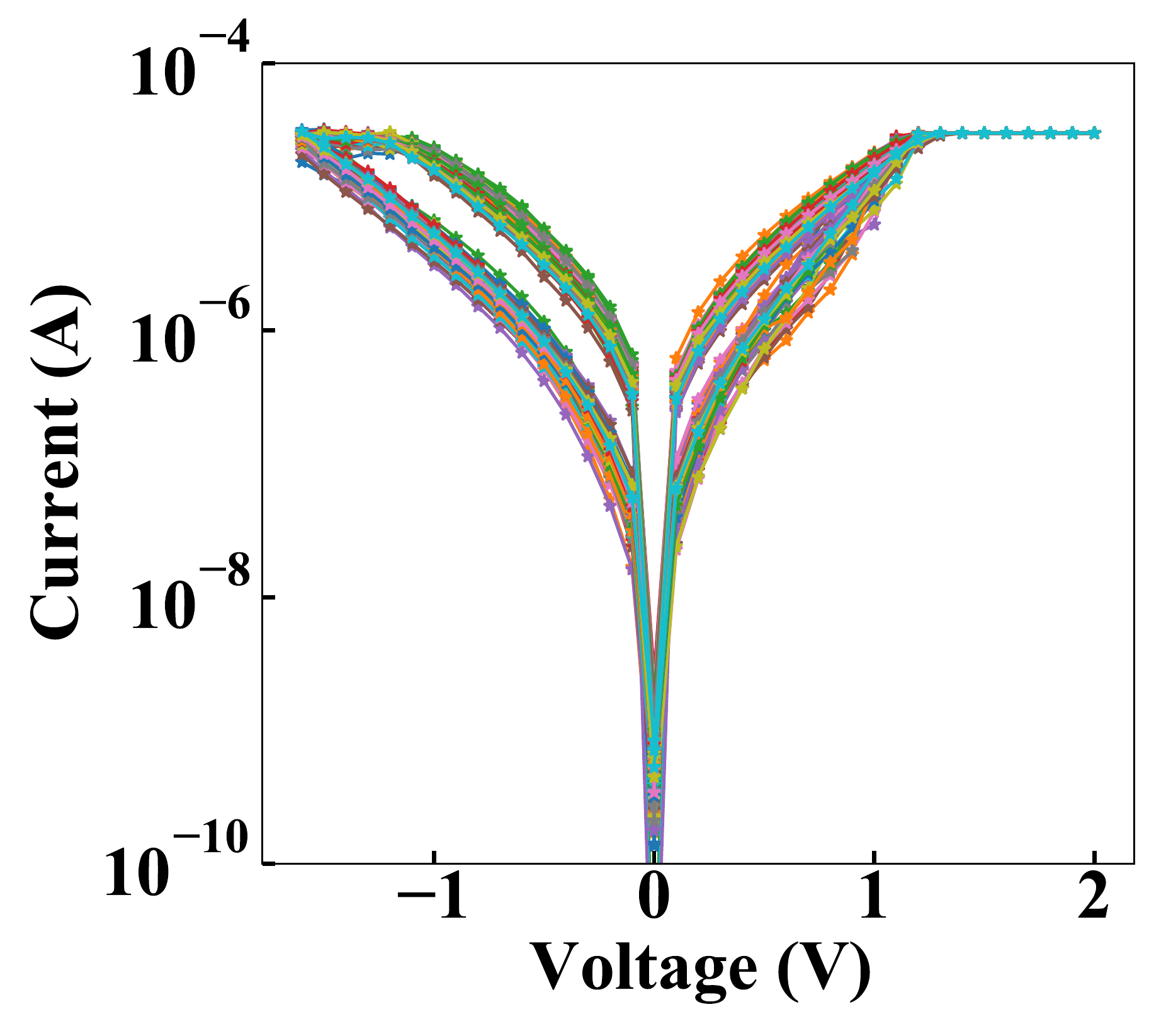}
	\caption{Current-Voltage ($I$-$V$) characteristics for cycle-to-cycle variation of the RRAM device.}
	\label{RRAM_C2C}
\end{figure}

\begin{figure}[!b]
	\subfigure[]{\includegraphics[scale=0.24]{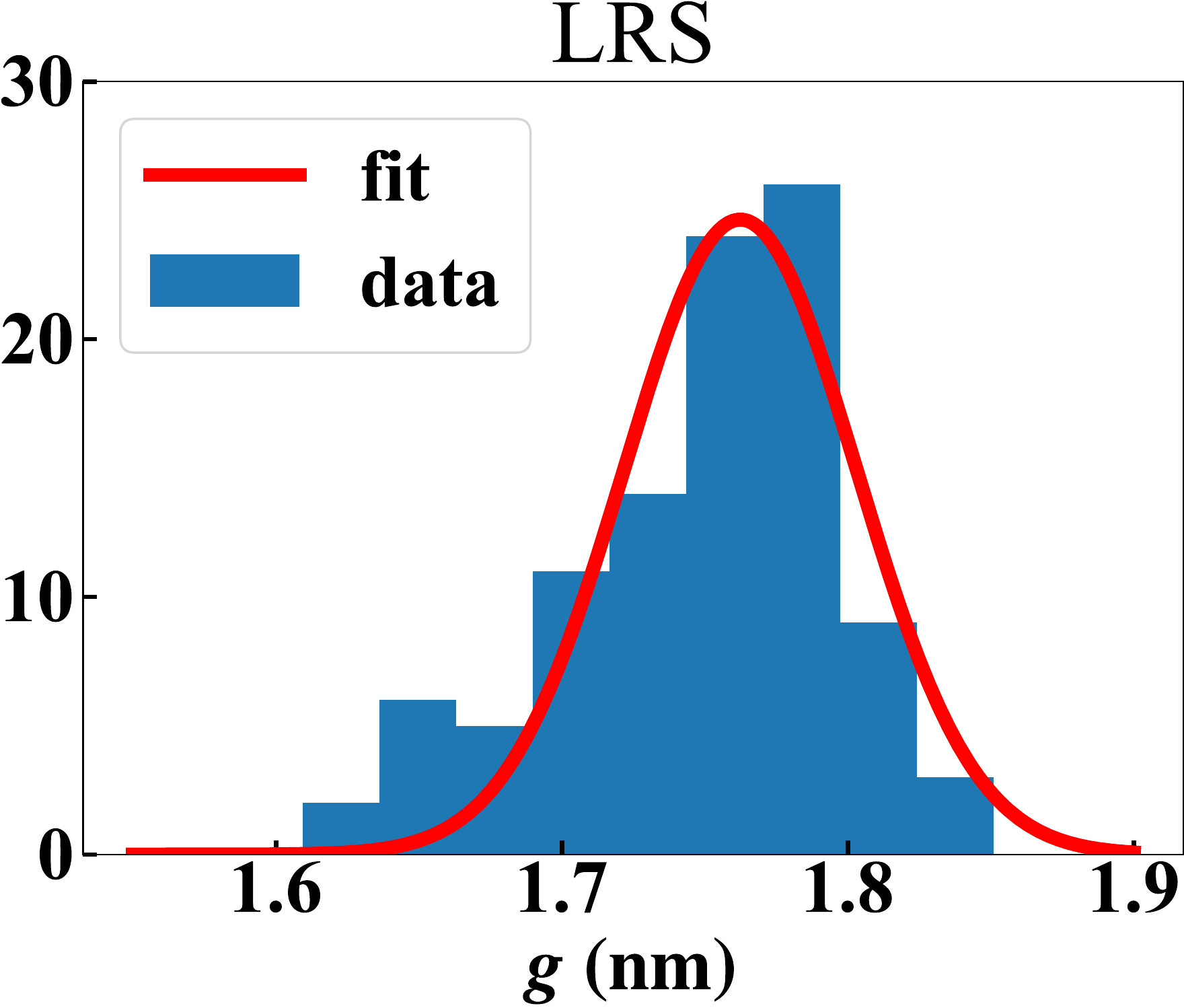}}
	\hfill
	\subfigure[]{\includegraphics[scale=0.24]{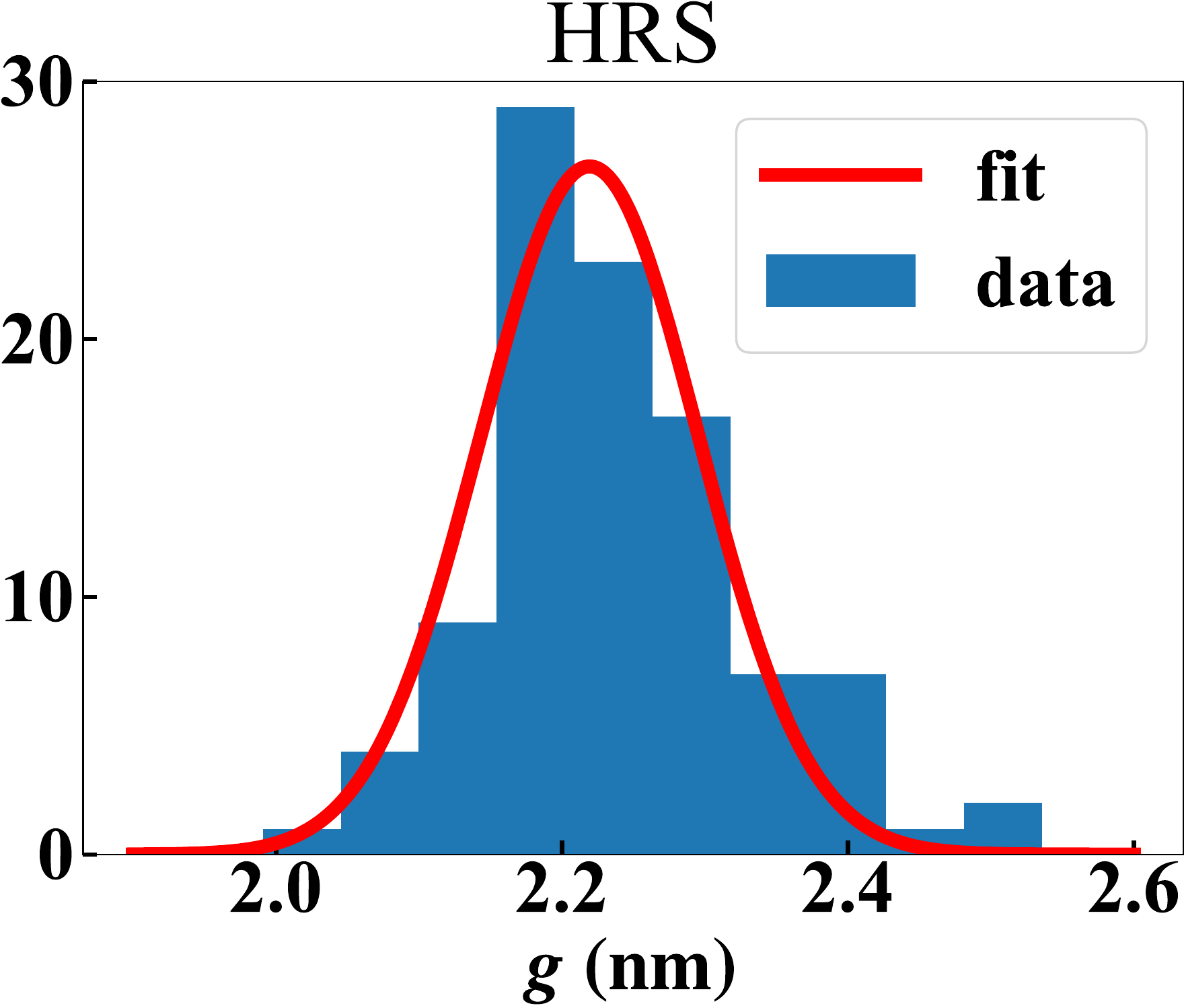}}
	\caption{Histogram and gaussian fit plot of the (a) LRS and (b) HRS gap lengths, of the 100 experimental data.}
	\label{LRS_HRS_gap_hist}
\end{figure}

As mentioned earlier, variation in filamentary RRAM resistance can be due to the stochastic nature of the gap length, $g$, which is reflected in the $I$-$V$ characteristics as shown in Fig.~\ref{RRAM_C2C}. Here, we plot the $I$-$V$ characteristics of cycle-to-cycle variation of RRAM device from experimental data of 100 cycles, where each color represents one cycle. The $I$-$V$ relationship of a RRAM device may be calculated using \cite{jiang2014verilog}
\begin{equation}
I=I_0~\text{exp}\left(-\frac{g}{g_0}\right)~\text{sinh}\left(\frac{V}{V_0}\right)
\label{RRAM_I_V}
\end{equation}where $I_0=0.25~\text{mA}$, $g_0=0.25~\text{nm}$, and $V_0=0.25~\text{V}$. We consider the RRAM to have two stable resistance states: the low resistance state (LRS) and the high resistance state (HRS). From Fig.~\ref{RRAM_C2C}, we calculate the LRS and HRS gap length at applied voltage of $-0.1~\text{V}$ using Eq.~(\ref{RRAM_I_V}) for 100 cycles. The histogram of the gap lengths is plotted in Fig.~\ref{LRS_HRS_gap_hist} and a Gaussian distribution having mean and standard deviation of 1.748~nm and 0.048~nm, respectively, is fitted to the histogram. For HRS, the Gaussian distribution that fits the histogram has mean and standard deviation of 2.236~nm and 0.095~nm respectively.

\begin{figure}[!b]
	\subfigure[]{\includegraphics[scale=0.2]{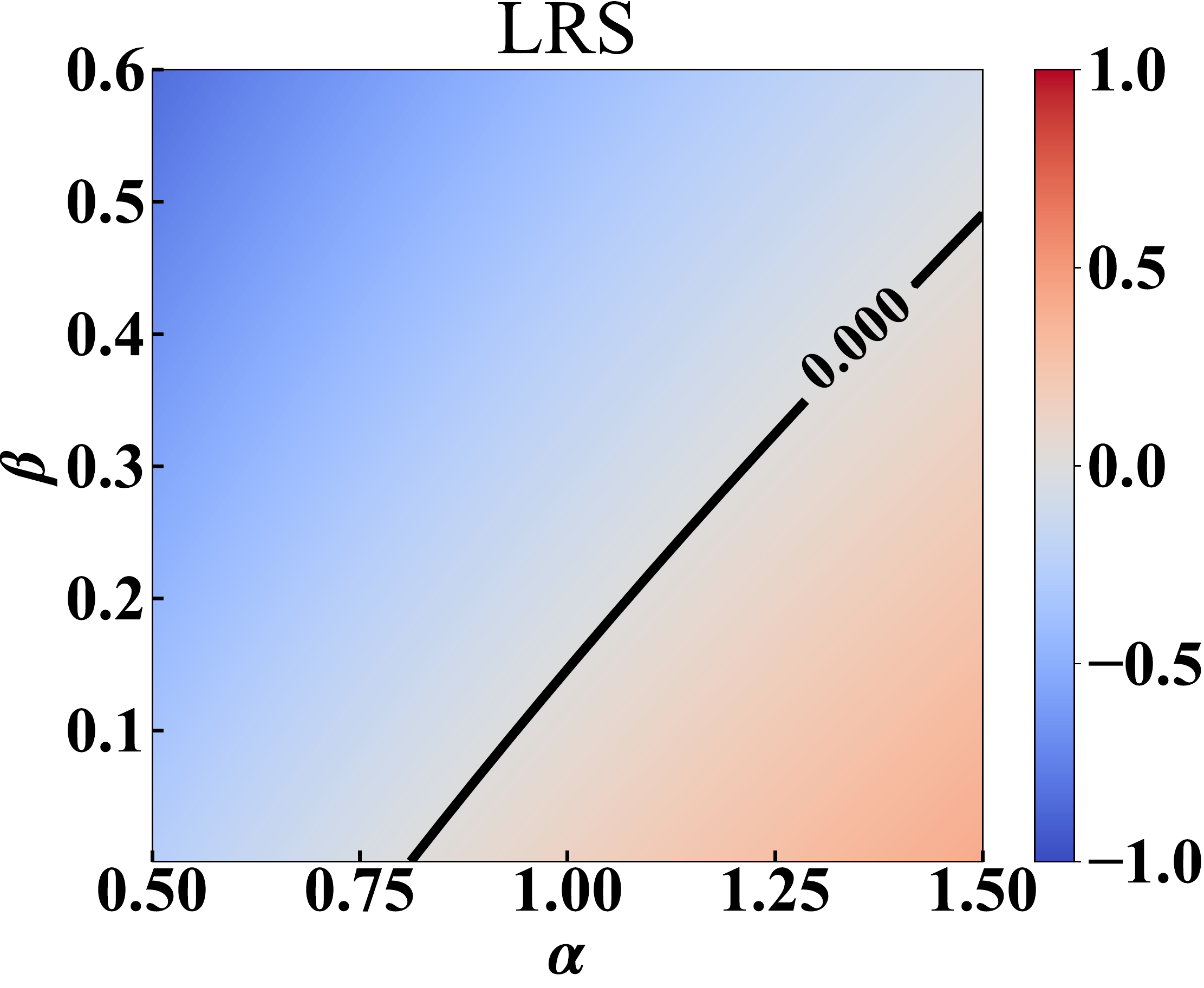}}
	\hfill
	\subfigure[]{\includegraphics[scale=0.2]{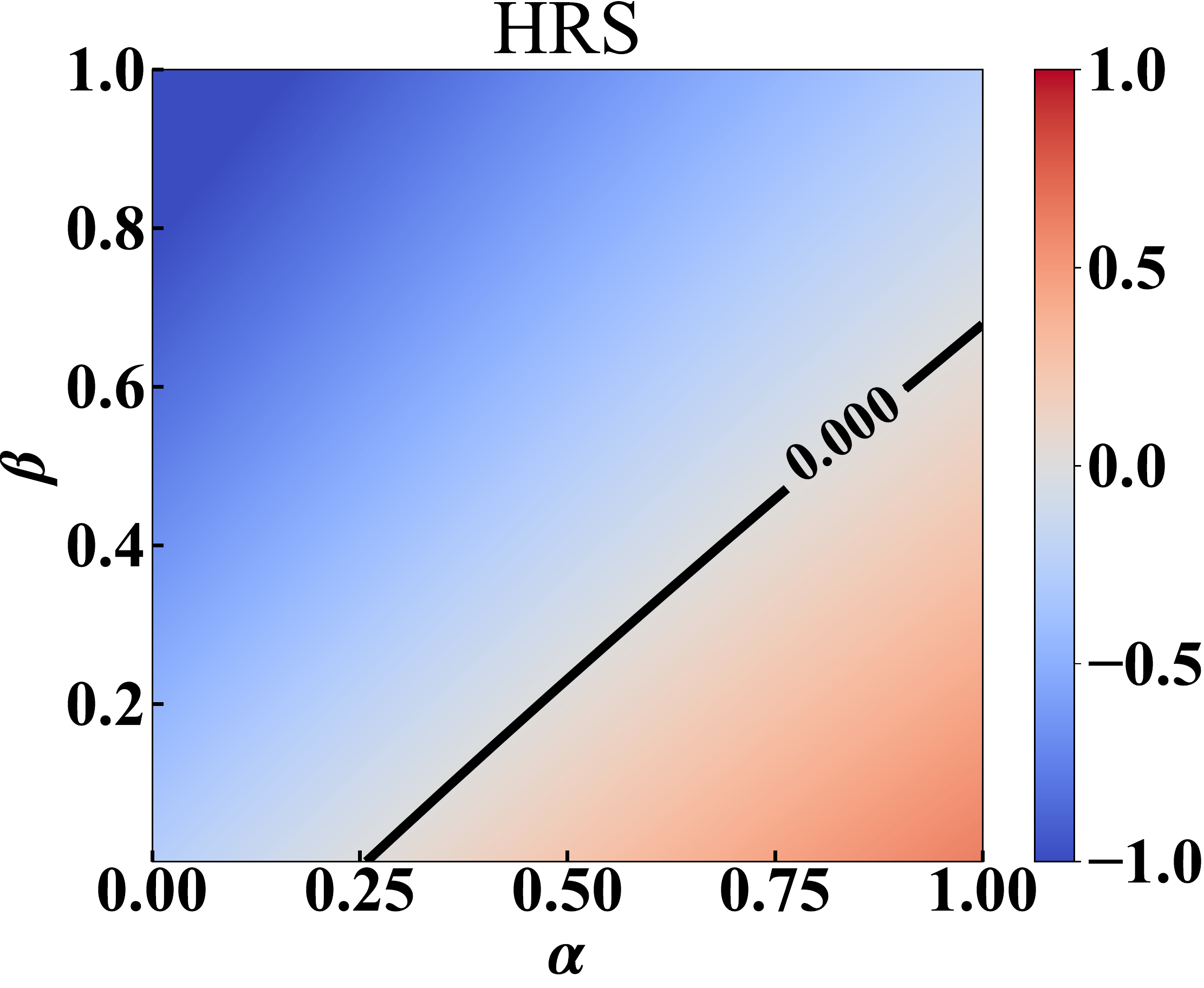}}
	\caption{Colormap of $h(\alpha, \beta)$ of Eq. (\ref{f_alpha_beta}), for (a) LRS and (b) HRS gap lengths. Black continuous line shows $h(\alpha, \beta)=0$. Colorbar represents the value of $h(\alpha, \beta)$.}
	\label{LRS_HRS_fmin}
\end{figure}

Since in our FP model, $f_\text{free}$ gives rise the distribution of $g$, we assume $f_\text{free}$ to have a minimum at the median of the gap length distribution. Furthermore, we consider $f_\text{free}$ to have the form
\begin{equation}
f_\text{free}(g,t) = A_0(t)~\text{exp}(-\alpha(t) g) + B_0(t)~\text{exp}(\beta(t)(g-g_0)) \label{RRAM_energy_barrier}
\end{equation}
where $A_0(t)$, $B_0(t)$, $\alpha(t)$, $\beta(t)$ are to be determined. In principle, time dependent effects in RRAM devices (e.g. resistance drift) may be modeled through the time dependence of $A_0(t)$, $B_0(t)$, $\alpha(t)$, $\beta(t)$. Also, $0<g<g_0$ where $g_0$ denotes the maximum value of the gap length for the corresponding distribution. Unlike MRAM, as discussed in the previous section, we explore the steady-state analysis of the RRAM device in this work. In this model, it is assumed that the stochastic change in gap length for the RRAM device arises during the SET and RESET processes. This can arise due to the random scattering of oxygen vacancies as they migrate. The time-scale of these scattering processes is much smaller than the time-scale of the SET and RESET processes. In our model, once the SET and RESET processes are complete, the distribution of the gap length is fixed. As our motivation is to demonstrate the promise of our FP based approach to model the LRS and HRS resistance distributions in the RRAM device, state transition from LRS to HRS and vice versa, which needs a detailed mathematical formulation, is beyond the scope of this paper and will be left as future work to be pursued. Since we are analyzing the steady state, the variable $t$ will be omitted from the mathematical formulations in rest of the manuscript. $A_0$ and $B_0$ are determined by considering $f_\text{free}(g=0) = f_\text{free}(g=g_0) = K = 1$. Then,
\begin{equation}
	A_0 + B_0~\text{exp}(-\beta g_0) = K \label{Cond_1}
\end{equation}\begin{equation}
	A_0~\text{exp}(-\alpha g_0) + B_0 = K \label{Cond_2}
\end{equation}
From Eqs.~(\ref{Cond_1}) and (\ref{Cond_2}), one obtains
\begin{IEEEeqnarray}{rCl}
A_0 & = & \frac{K\left[1-\text{exp}(-\beta x_0)\right]}{1-\text{exp}\left[-(\alpha+\beta)x_0\right]}\\
B_0 & = & \frac{K\left[1-\text{exp}(-\alpha x_0)\right]}{1-\text{exp}\left[-(\alpha+\beta)x_0\right]} \label{A0B0}
\end{IEEEeqnarray}
Since $\frac{d}{dg}f_\text{free}|_{g=g_\text{min}} = 0$ if $f_\text{free}$ has a minimum at $g=g_\text{min}$, differentiating Eq.~(\ref{RRAM_energy_barrier}) with respect to $g$ and setting it to zero at $g=g_\text{min}$ gives
\begin{equation}
	-\alpha~A_0~\text{exp}(-\alpha g_\text{min}) + \beta~B_0~\text{exp}(\beta(g_\text{min}-g_0)) = 0 \label{dEdg_eq_0}
\end{equation}
From Eq.~(\ref{dEdg_eq_0}), $g_\text{min}$ is given by the expression 
\begin{equation}
	g_\text{min} = \frac{1}{(\alpha+\beta)}\left\lbrace ln\left[\left(\frac{\alpha}{\beta}\right)\frac{1-\text{exp}(-\beta g_0)}{1-\text{exp}(-\alpha g_0)}\right]+\beta g_0\right\rbrace \label{g_min}
\end{equation}
To find $\alpha$ and $\beta$, Eq. (\ref{g_min}) is first rewitten as 
\begin{align}
	\begin{split}
		& h(\alpha, \beta) \\ &=  g_\text{min} - \frac{1}{(\alpha+\beta)} \left\lbrace \text{ln}\left[\left(\frac{\alpha}{\beta}\right)\frac{1-\text{exp}(-\beta g_0)}{1-\text{exp}(-\alpha g_0)}\right] +\beta g_0 \right\rbrace\\ 
	\end{split} \label{f_alpha_beta}
\end{align}
For given values of $g_0$ and $g_\text{min}$, $\alpha$, $\beta$ are varied to find the $h(\alpha, \beta)$=0 curve using the contour plot shown in Fig.~\ref{LRS_HRS_fmin} by the black continuous line. From this plot, the allowed range of $\alpha$ and $\beta$ for both LRS and HRS gap lengths can be found. $\beta$ can be numerically calculated by first choosing a value of $\alpha$ and numerically solving Eq.~(\ref{f_alpha_beta}) for $h(\alpha, \beta)=0$ and with a known value of $g_\text{min}$. The obtained values of $\alpha$, $\beta$ and $K$, are then used to calculate $A_0$ and $B_0$ using Eq.~(\ref{A0B0}) (see Table \ref{Table_2}). Finally, $f_\text{free}$ for LRS and HRS are obtained using Eq.~(\ref{RRAM_energy_barrier}).

\begin{table}[h]
	\centering
	\caption{\label{Table_2}Parameter values of LRS and HRS}
	\begin{tabular}{ |c|c|c|c|c| } 
		\hline
		& $\alpha$ & $\beta$ & $A_0$ & $B_0$ \\
		\hline 
		LRS & 0.9 & 0.07104 & 0.30130 & 0.9966\\ 
		\hline
		HRS & 0.27 & 0.01288 &0.08241 & 0.97846\\
		\hline		
	\end{tabular}
\end{table}

\begin{figure}[!t]
	\subfigure[]{\includegraphics[scale=0.24]{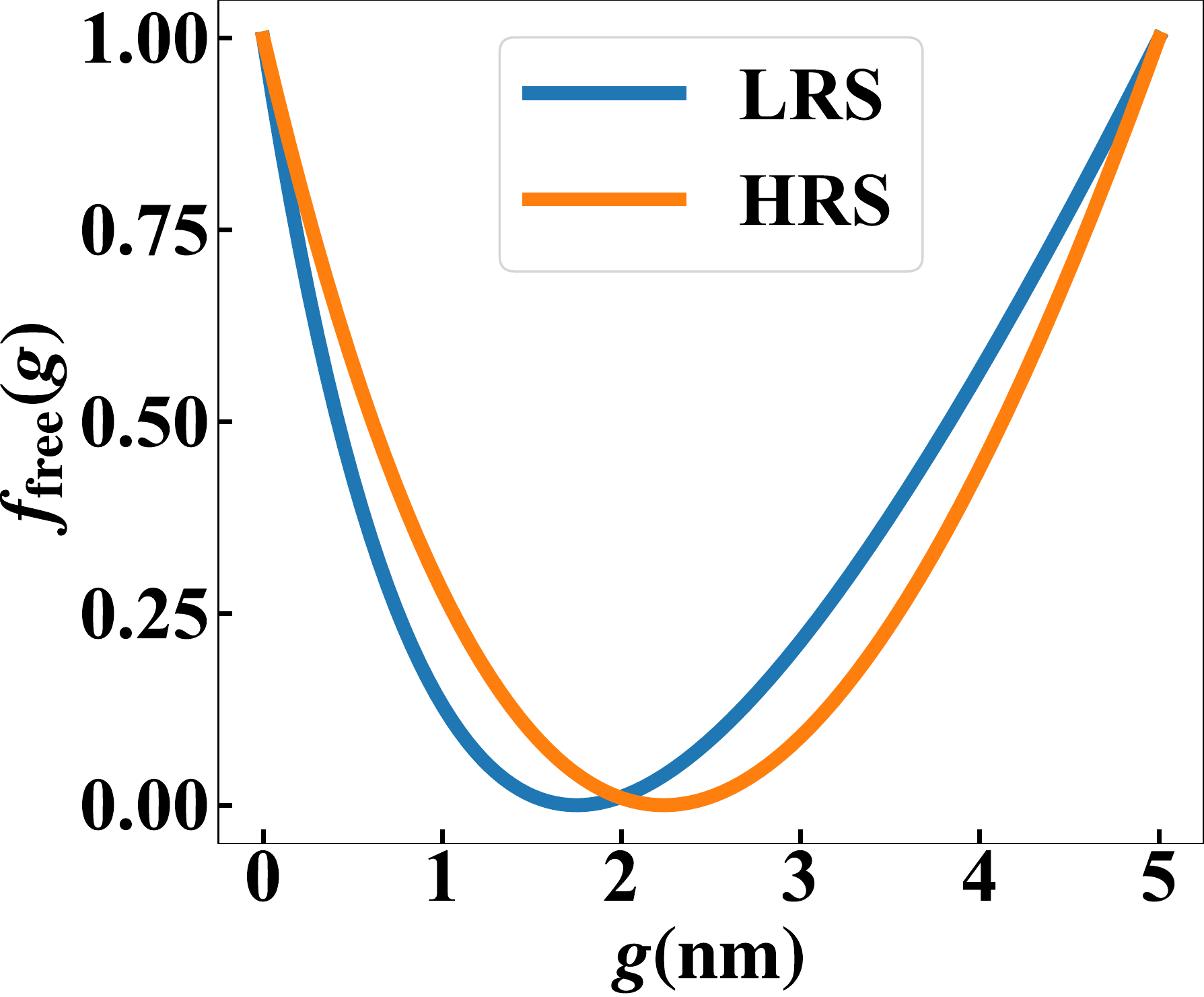}}
	\hfill
	\subfigure[]{\includegraphics[scale=0.24]{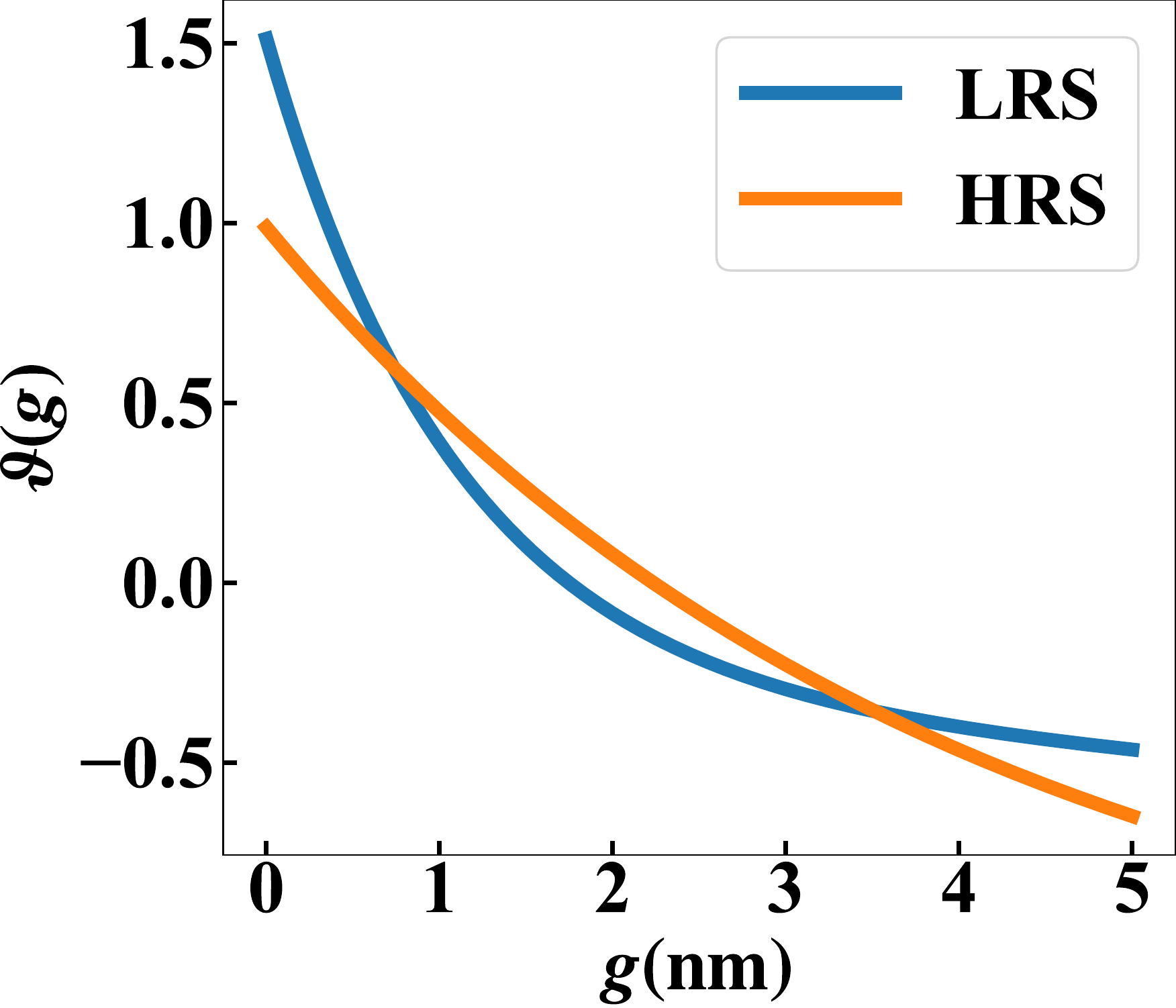}}
	
	\caption{(a) Profile of the free energy function $f_{\text{free}}(g)$ for LRS and HRS and (b) the corresponding velocity profile $\vartheta(g)$ for LRS and HRS, respectively.}
	\label{Egvg}
\end{figure}
\begin{figure}[!t]
	\subfigure[]{\includegraphics[scale=0.24]{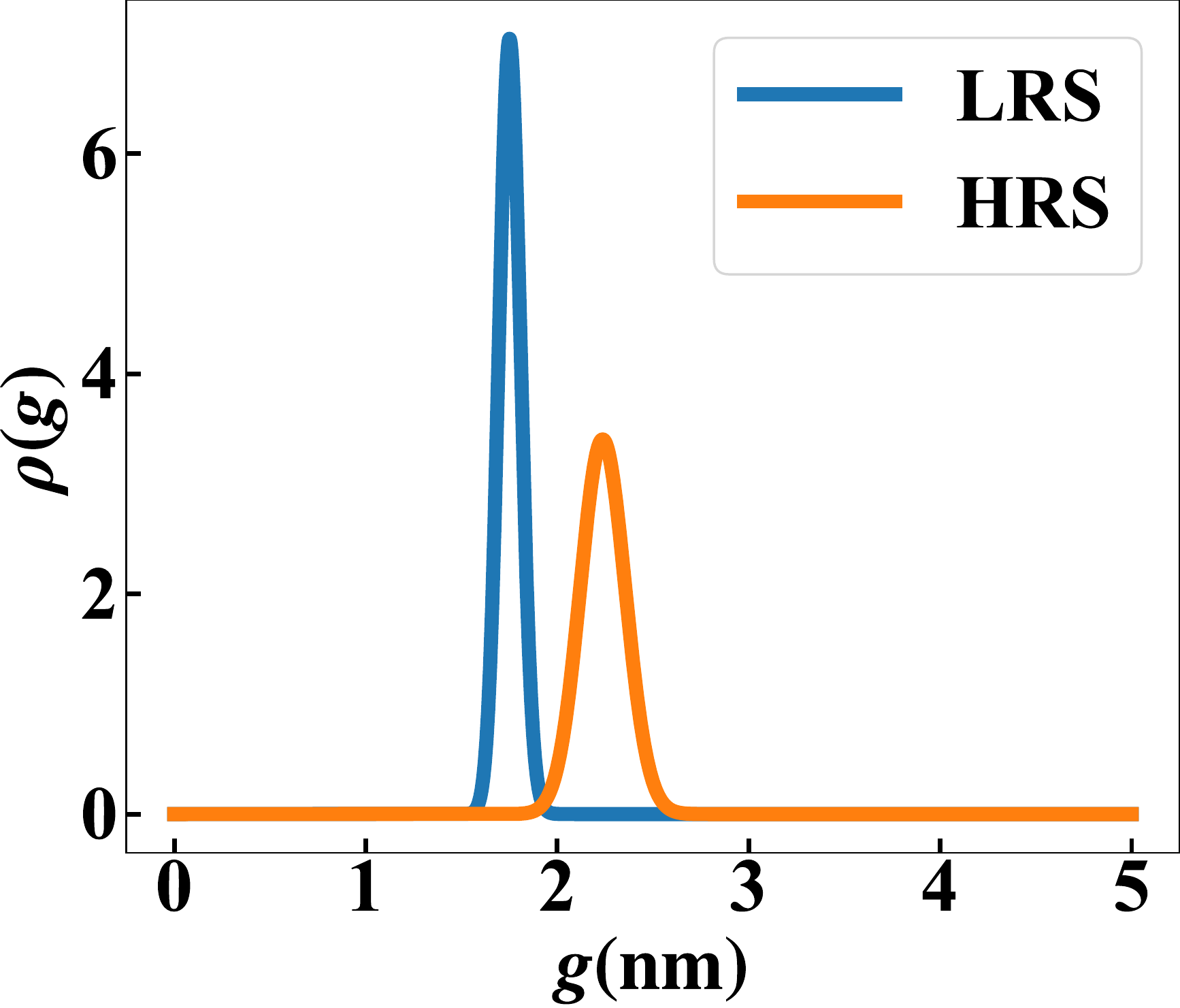}}
	\hfill
	\subfigure[]{\includegraphics[scale=0.24]{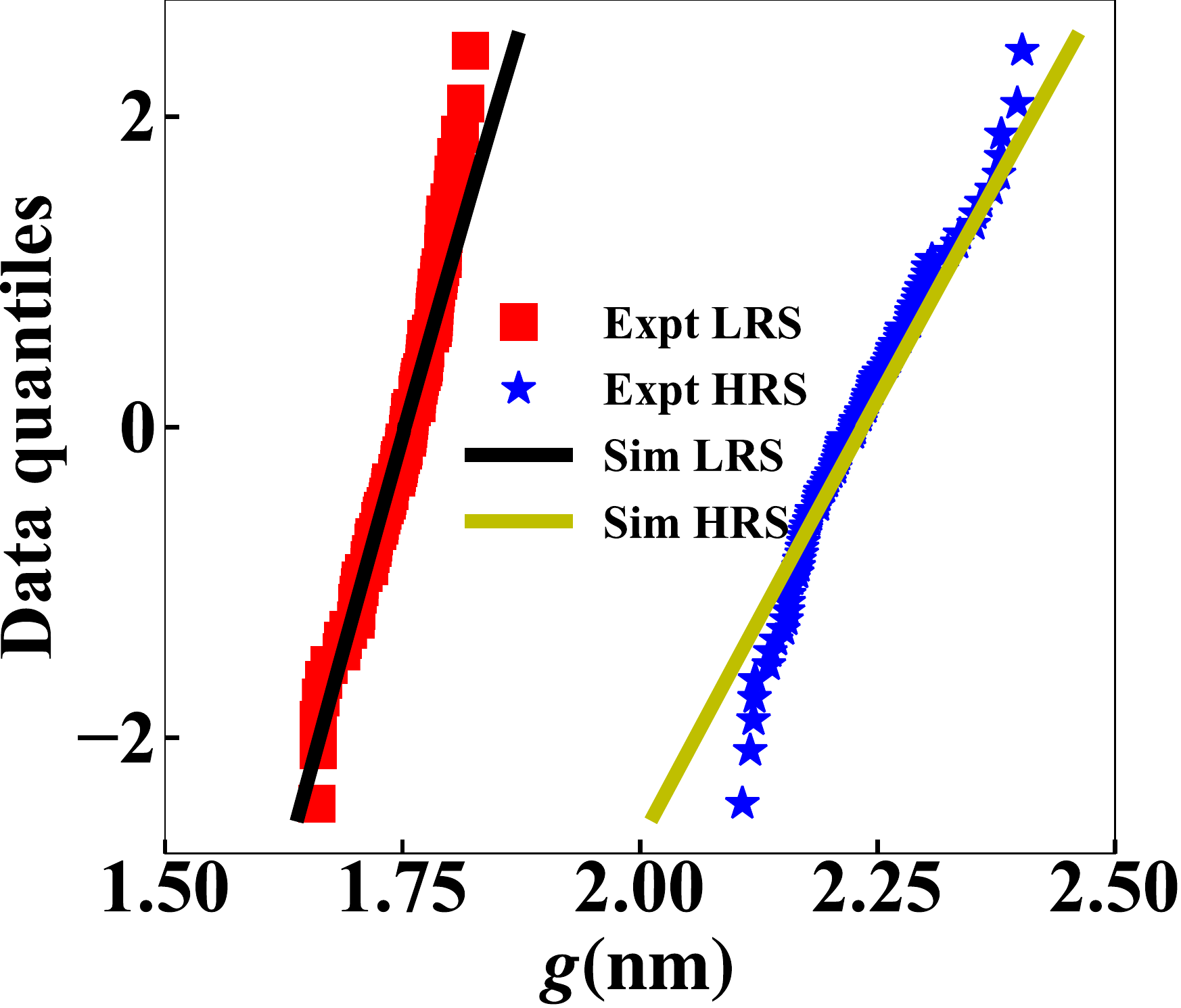}}
	
	\caption{(a) Probability distribution function (PDF) of LRS and HRS, obtained from the FP simulation. (b) Comparison of the quantiles for the LRS and HRS gap length distribution, obtained from experiment and simulation.}
	\label{PDF_CDF}
\end{figure}

Based on the parameters in Table \ref{Table_2}, the corresponding velocity profile $\boldsymbol{\vartheta}(g)=-\frac{\partial f_\text{free}(g)}{\partial g}$ are plotted in Fig.~\ref{Egvg}. For the simulation, $g$ is chosen to be in the range 0-5 nm with 10000 dicrete points. Then, $D=\sigma^2/2$ in Eq.~(\ref{FP}), where $\sigma$ is the standard deviation of gap length distribution determined from Fig.~\ref{LRS_HRS_gap_hist}. The results after solving Eq.~(\ref{FP}) for $\rho(g)$ are plotted in Fig.~\ref{PDF_CDF}(a). To compare our simulation result with the experiment, the quantiles of the gap length distribution for LRS and HRS are plotted in Fig.~\ref{PDF_CDF}(b), where the symbols represent the experimental data and the solid lines represent data quantile for the gap length distribution returned by our simulation. First, the experimental data quantile was plotted and found it was found to be limited to $\sim\pm 2.5$. Next, the quantiles of our simulation result the quantile within $\pm 2.5$ was extracted for a fair comparison. Fig.~\ref{PDF_CDF}(b) shows that our simulation result matches well with the experimental results and confirms the accuracy of our simulation methodology.

\begin{figure}[!t]
	\includegraphics[scale=0.22]{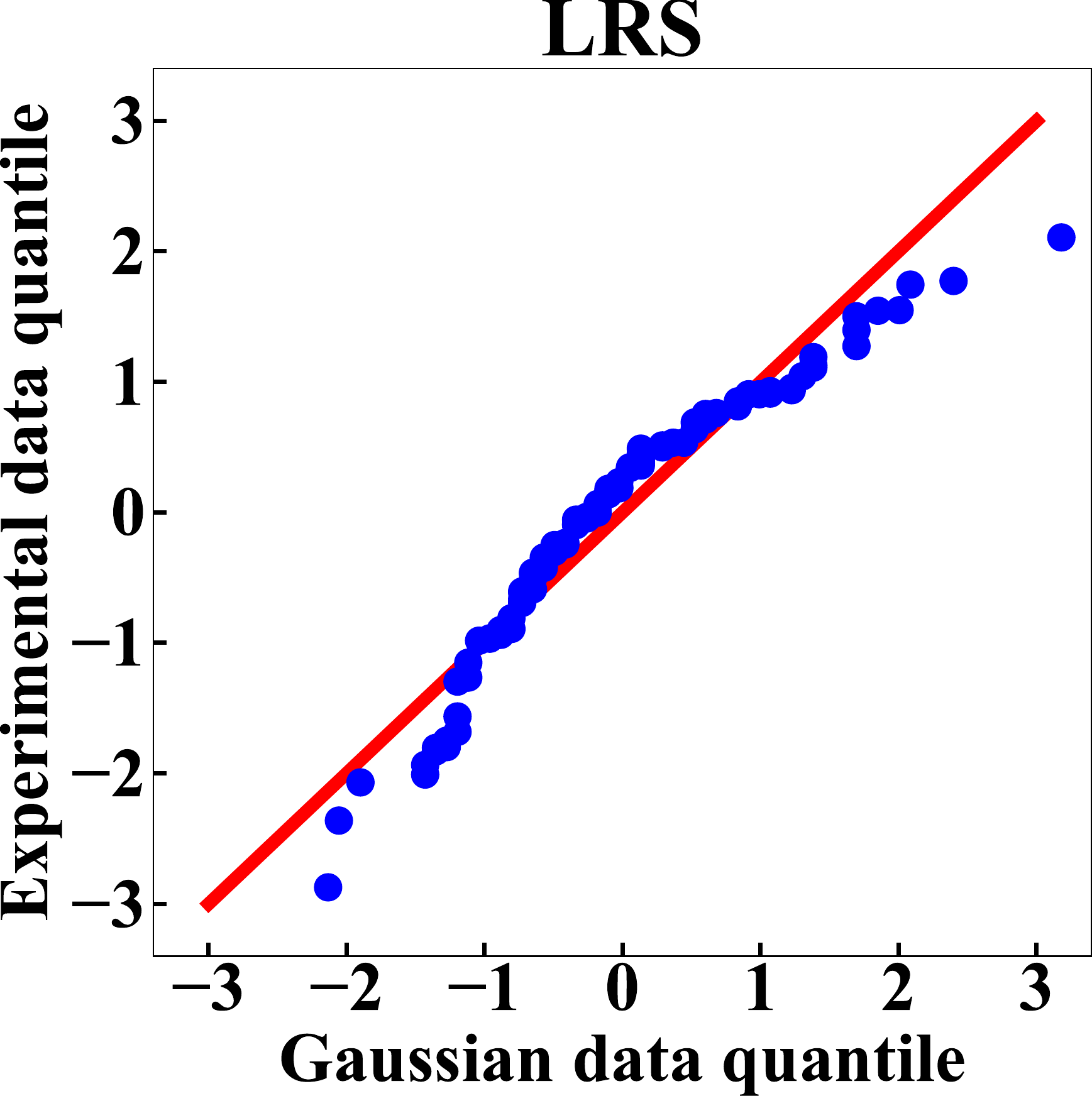}
	\hfill
	\includegraphics[scale=0.22]{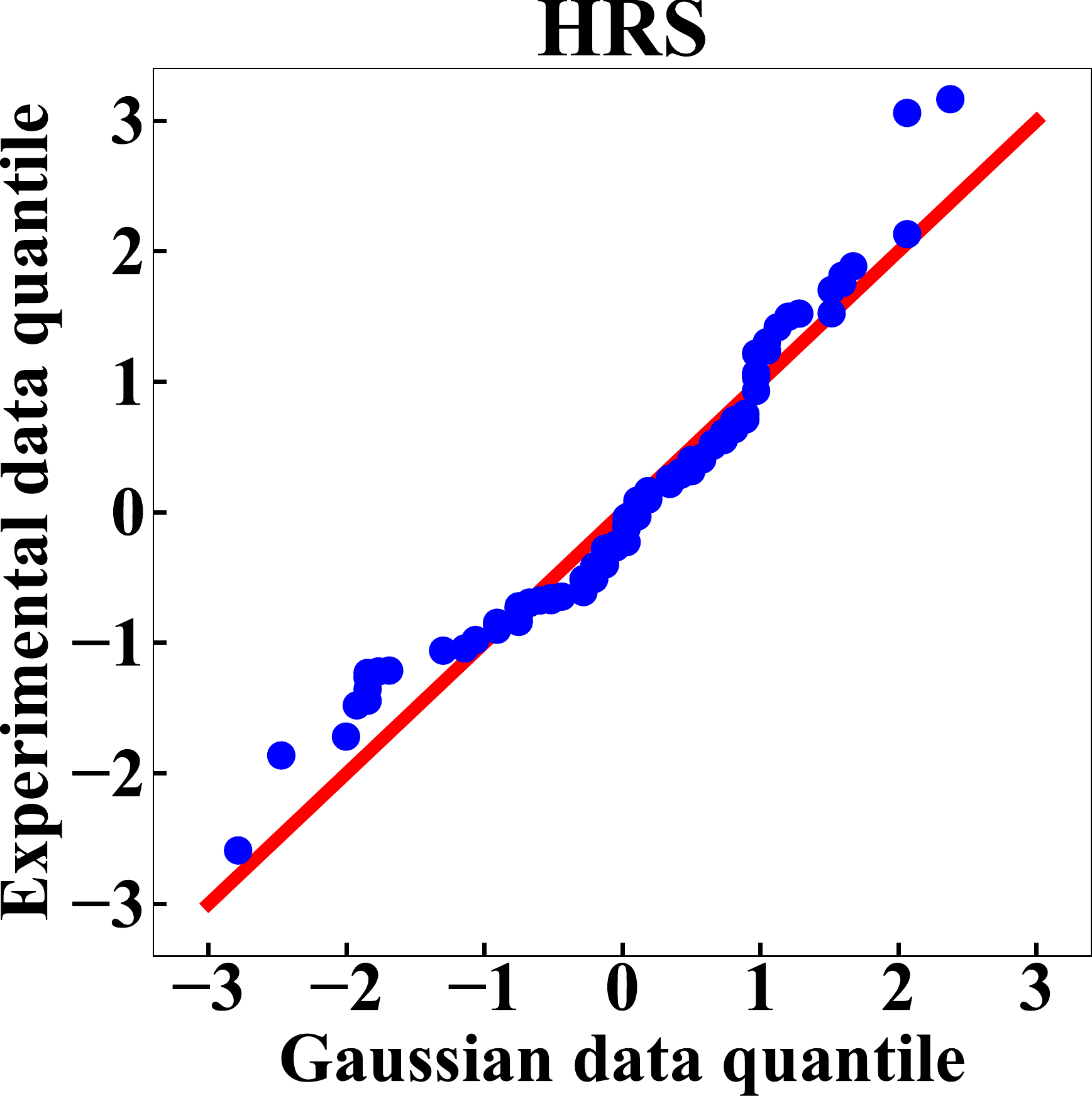}
	\begin{center}
		(a)
	\end{center}
	\includegraphics[scale=0.22]{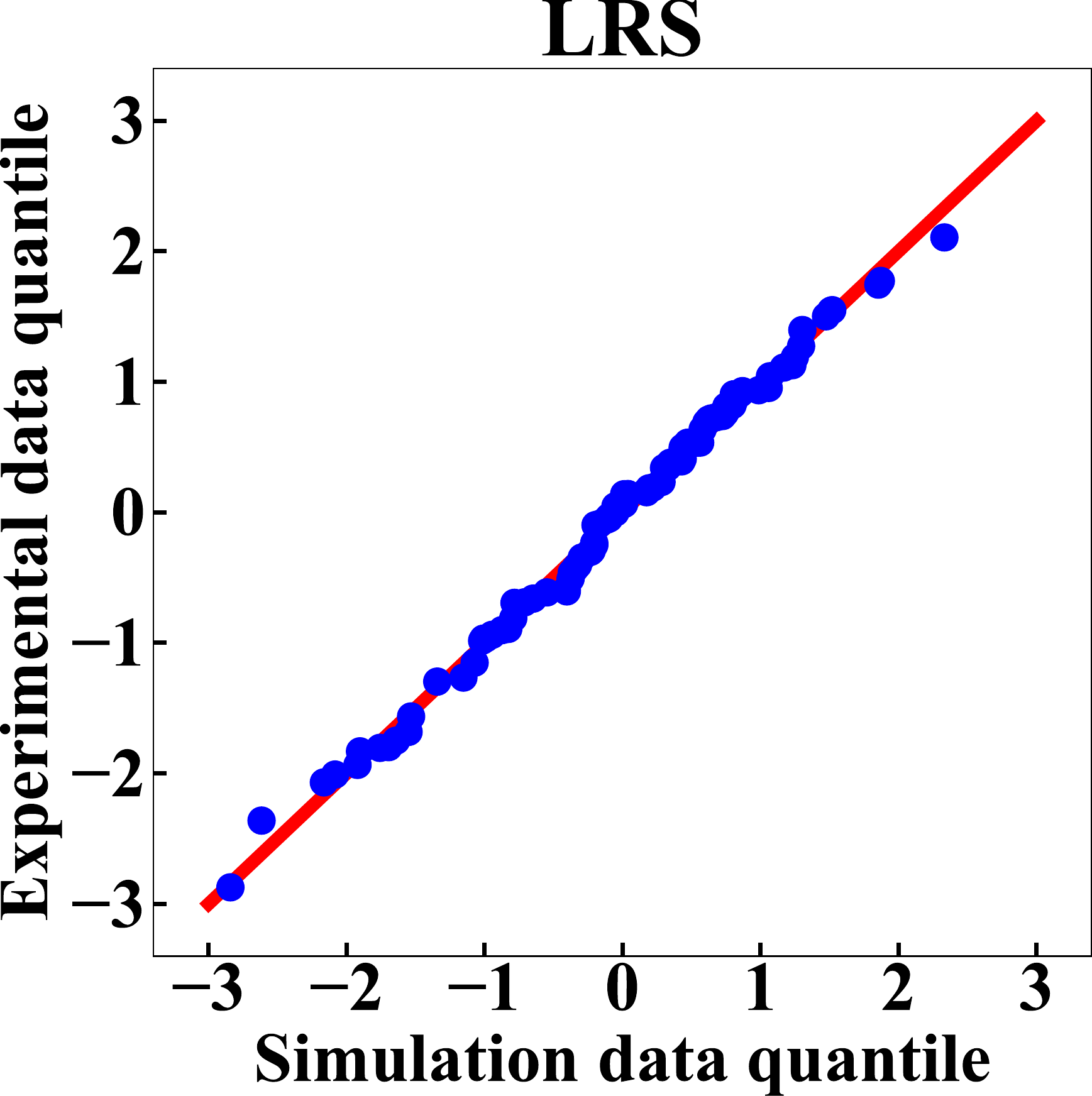}
	\hfill
	\includegraphics[scale=0.22]{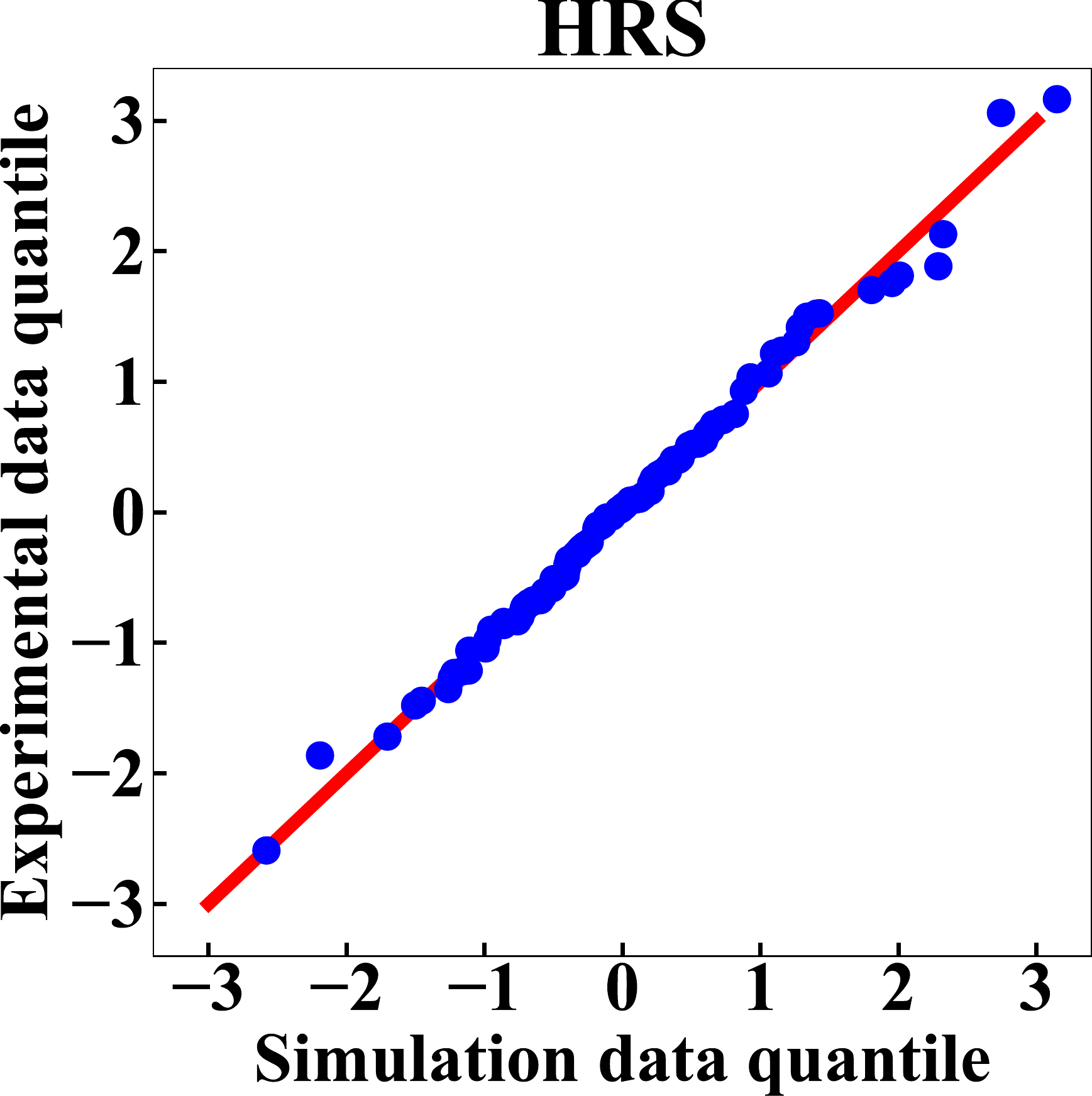}
	\begin{center}
		(b)
	\end{center}
	\caption{(a) QQ plot between experimental and Gaussian distribution quantiles of LRS and HRS gap lengths. (b) QQ plot between experimental and simulation quantiles of LRS and HRS gap lengths. Blue dots represents quantile whereas red straight line denotes $y=x$ relation between experimental and Gaussian distribution(simulation) quantile data. }
	\label{qqplot}
\end{figure}

Finally, quantile-quantile (QQ) plots are generated to compare the quantiles of the experimental data to the quantiles of the gap length distribution obtained from our simulation (for both LRS and HRS), as shown in Fig.~\ref{qqplot}. QQ-plots are preferred because errors in fitting the tails of the distribution can be exposed. If the data quantiles matches the quantiles of the assumed distribution, the data will lie along a straight line through the origin. Deviation of the slope from $+1$ indicates that the distribution of the data is the same as the distribution it is compared to except with a different standard deviation. First, the quantiles of the experimental gap length distribution for LRS and HRS are compared to the Gaussian distributions in Fig.~\ref{LRS_HRS_gap_hist}. The QQ-plots for LRS and HRS are shown in Fig.~\ref{qqplot}(a). The data quantiles are plotted by blue circles. A straight line with $y=x$ (in red color) is added to aid in comparing the distribution of the data with the Gaussian distributions. The graph shows that the data quantiles do not match well to the Gaussian distribution. Next, the experimental quantiles are compared with those obtained from our FP-based model in Fig.~\ref{qqplot}(b). From this plot, we see that the distribution obtained by our simulation better fits the distribution of the experimental data. This is also indicated by the mean square error (MSE) for the QQ plots. For fitting the experimental data to Gaussian distribution in Fig. \ref{qqplot}(a), the MSEs are 0.078 and 0.064 for LRS and HRS, respectively. For Fig. \ref{qqplot}(b) (i.e. fitting experimental data to our FP simulation), the MSE values are 0.005 and 0.007, respectively. Hence, we conclude that our proposed FP approach is able to replicate the distribution of the gap length in the experimentally measured RRAM data.

\section{Conclusion}\label{conclusion}

We presented an approach based on the FP equation to capture the impact of stochastic programming process in RRAM and MTJ devices as an alternative to the computationally expensive Monte-Carlo approach. The WER of the MTJ was investigated using our simulation approach and verified with the experimentally reported data. The effect of the applied voltage across the MTJ and the VCMA effect on the WER are also discussed. The effect of dipolar coupling on WER in MRAM is also explored using the framework. We then showed that our FP approach can also reproduce the LRS and HRS resistance distributions in RRAM. The mathematical formulation and methodology to model the distribution of the gap length of conductive filament in RRAM were discussed. Finally, we showed that the distribution of the gap length obtained from our simulation matches well with the experimental data.

\section{acknowledgement}
This research project is supported in part by the National Research Foundation, Singapore, under its Competitive Research Programme (Award No. NRF-CRP24-2020-0003), in part by the A*STAR SpOT-LITE Programmatic Grant, and the Ministry of Education (Singapore) Tier~1 Academic Research Fund. The authors would like to thank Dr. Subhranu Samanta for providing the unpublished experimental data for the RRAM device.

\bibliographystyle{IEEEtran}
\bibliography{References}
\end{document}